\documentclass[11pt]{article}

\usepackage{jheppub}
\usepackage{amsmath}
\usepackage{mathrsfs}
\usepackage{bbm}
\usepackage{amsfonts}
\usepackage{amssymb}
\usepackage{latexsym}
\usepackage{graphicx}
\usepackage[english]{babel}
\usepackage{multirow}
\usepackage{float}
\usepackage{url}
\usepackage{slashed}
\usepackage{tabularx}
\usepackage{orcidlink}
\usepackage{appendix}
\usepackage{physics}
\usepackage[utf8]{inputenc}

\newcommand{\be}{\begin{equation}}
\newcommand{\ee}{\end{equation}}
\newcommand{\ba}{\begin{array}}
\newcommand{\ea}{\end{array}}
\newcommand{\bea}{\begin{eqnarray}}
\newcommand{\eea}{\end{eqnarray}}

\title{Refining Gravitational Wave and Collider Physics Dialogue via Singlet Scalar Extension}

\author[a,b,c,d]{Michael J.~Ramsey-Musolf,}
\author[a,e]{Tuomas V.~I.~Tenkanen}
\author[a,f,g]{and Van Que Tran} 

\preprint{HIP-2024-20/TH}

\affiliation[a]{Tsung-Dao Lee Institute \& School of Physics and Astronomy, Shanghai Jiao Tong University, Shanghai 200240, China}
\affiliation[b]{Shanghai Key Laboratory for Particle Physics and Cosmology, Key Laboratory for Particle Astrophysics and Cosmology (MOE), Shanghai Jiao Tong University, Shanghai 200240, China}
\affiliation[c]{Amherst Center for Fundamental Interactions, Department of Physics, University of Massachusetts, Amherst, MA 01003, USA}
\affiliation[d]{Kellogg Radiation Laboratory, California Institute of Technology, Pasadena, CA 91125, USA}
\affiliation[e]{Department of Physics and Helsinki Institute of Physics,
    P.O.~Box 64, FI-00014 University of Helsinki,
    Finland}
    
\affiliation[f]{Institute of Physics, Academia Sinica, Nangang, Taipei 11529, Taiwan}

\affiliation[g]{Phenikaa Institute for Advanced Study, Phenikaa University, Yen Nghia, Ha Dong, Hanoi 100000, Vietnam}

\emailAdd{mjrm@sjtu.edu.cn}
\emailAdd{tuomas.tenkanen@helsinki.fi}
\emailAdd{vqtran@sjtu.edu.cn}

\abstract{
Employing effective field theory techniques, we advance computations of thermal parameters that enter predictions for the gravitational wave spectra from first-order electroweak phase transitions. Working with the real-singlet-extended Standard Model, we utilize recent lattice simulations to confirm the existence of first-order phase transitions across the free parameter space. For the first time, we account for several important two-loop corrections in the high-temperature expansion for determining thermal parameters, including the bubble wall velocity in the local thermal equilibrium approximation. 
We find that the requirement of completing bubble nucleation  
imposes stringent bounds on the new scalar boson mass.
Moreover, the prospects for detection by LISA require first-order phase transitions in a two-step phase transition, which display strong sensitivity to the portal coupling between the Higgs and the singlet. 
Interestingly, signals from di-Higgs boson production at the HL-LHC probe parameter regions that significantly overlap with the LISA-sensitive region, indicating the possibility of accounting for both signals if detected. Conversely, depending on the mixing angle, a null result for di-Higgs production at the HL-LHC could potentially rule out the model as an explanation for gravitational wave observations.
}

\begin{document}
\maketitle

%%%%%%%%%%%%%%%%%%%%%%%%%%%%% SECTION %%%%%%%%%%%%%%%%%%%%%%%%%%%%%%%%%%%%
%

\section{Introduction}
\label{sec:intro}

Unravelling the thermal history of the electroweak symmetry breaking
remains as a fascinating challenge in the intersection of particle physics and cosmology.
Lattice simulations have revealed that as the temperature drops below the electroweak scale ($\sim$ 100 GeV) the minimal Standard Model (SM) smoothly transitions from the deconfinement phase to the Higgs phase through a crossover \cite{Kajantie:1996mn,Csikor:1998eu}. However, extending the scalar sector of the Standard Model can  lead to a first-order electroweak phase transition (EWPT), which could provide an out-of-the-equilibrium 
Sakharov condition for the generation of the matter-antimatter asymmetry through the electroweak baryogenesis \cite{Kuzmin:1985mm,Morrissey:2012db,Bodeker:2020ghk}. 
Mapping out the phase diagrams of interesting Standard Model extensions has been a major goal of the multiple decade lasting, still on-going program aiming to understand cosmological ramifications of these theories during the electroweak epoch after the Hot Big Bang. 

If the EWPT is strong enough, and completes through nucleation of bubbles of the Higgs phase, this violent early universe process generates sound waves \cite{Hindmarsh:2013xza,Hindmarsh:2015qta,Cutting:2019zws} and turbulence \cite{Dahl:2021wyk,Auclair:2022jod,Dahl:2024eup} in the hot plasma that can distort the spacetime leading to a production of a stochastic background of gravitational waves (GW). For phase transitions at temperatures around the electroweak scale, such GW signals are produced at milli-Hertz range and can be probed by future space-based interferometers such as 
LISA \cite{Audley:2017drz,Cornish:2018dyw}, 
DECIGO \cite{Kudoh:2005as,Kawamura:2011zz, Musha:2017usi}, 
BBO \cite{Crowder:2005nr, Yagi:2011wg}, 
TAIJI \cite{Gong:2014mca, Hu:2017mde, Guo:2018npi}, and 
TIANQIN \cite{TianQin:2015yph, Hu:2017yoc}, 
hence providing a remarkable new window to around one picosecond old universe.
For a review on the GW signatures from a first-order cosmological phase transition, see {\it e.g.}~\cite{Weir:2017wfa,Caprini:2019egz,LISACosmologyWorkingGroup:2022jok, Athron:2023xlk, Caprini:2024hue},
and \cite{Gowling:2021gcy,Gowling:2022pzb,Caprini:2024hue,Hindmarsh:2024ttn,Liang:2024tgn} for studies on their detection prospects.
Crucially, the mass scale of new scalar fields associated with interactions driving an EWPT cannot be too heavy with respect to the electroweak scale \cite{Ramsey-Musolf:2019lsf}. This presents an opportunity for new physics to be searched for at the LHC and other future colliders, e.g.~ILC \cite{ILC:2013jhg}, CEPC \cite{CEPCStudyGroup:2018ghi}, FCC \cite{FCC:2018byv} and CLIC \cite{CLIC:2018fvx}. 
In this work at hand, we concretely
explore the (non-$Z_2$ symmetric) scalar singlet extension of the Standard Model (``xSM'' \cite{Profumo:2007wc}), characterized by a rich collider phenomenology such as the high mass resonance di-Higgs production at the HL-LHC \cite{Apollinari:2017lan}.  

The methodology we use to study the early universe thermodynamics of the xSM, is generic to a wide range of theories beyond the Standard Model (BSM). 
The EWPT in many BSM theories has been extensively studied in the literature (see e.g.~\cite{Caprini:2019egz,Athron:2023xlk}) using perturbation theory, but  
a precise determination of
character of a phase transition -- i.e. whether transition is of first- or second-order, or a crossover (in which case there is no phase transition at all) --
requires non-perturbative methods. 
In perturbation theory, the derivative of the free-energy of the system with respect to temperature exhibits a discontinuity between different phases. In other words, the local minima of the thermal effective potential are separated by a potential barrier. Such barrier can be radiatively generated by loops of gauge bosons, or BSM scalars, or it can be present already at tree-level for a multi-field scalar potential. In particular, weak vector bosons induce such a barrier at one-loop at high temperatures through a cubic term in thermal effective potential. Hence, perturbation theory often predicts a first-order phase transition by default. 

This naive picture, however, can be misleading if transition is very weak and non-perturbative effects at high temperatures are significant. 
Lattice simulations \cite{Farakos:1994xh,Kajantie:1996mn,Csikor:1998eu} can be used to account for complicated, non-perturbative phenomena related to the symmetric phase, and indeed can find that there is no phase transition at all: for a crossover, all temperature derivatives of the free-energy, or the pressure, are continuous \cite{Kajantie:1995kf}. This is the case in the minimal Standard Model with Higgs masses $m_H \gtrsim m_W$ \cite{Kajantie:1996mn}.  

Phase diagram of a given model
can most readily be studied in
terms of an effective field theory (EFT) constructed using 
the high-temperature dimensional reduction \cite{Ginsparg:1980ef,Appelquist:1981vg}:
in this approach properties of the phase transition at long distances (at IR) are described by a static, three-dimensional effective theory.
Parameters of such thermal EFT capture temperature dependence of the full parent theory, and systematically include {\em thermal resummations} from short-distance, ultraviolet (UV) physics. For constructing thermal EFTs, see \cite{Kajantie:1995dw,Braaten:1995cm,Ekstedt:2022bff}.

Thermodynamic properties of such EFTs can be computed in terms of perturbation theory, c.f.~\cite{Arnold:1992rz, Farakos:1994kx,Laine:1994zq,Rajantie:1996np,Bodeker:1996pc,Gould:2021dzl,Ekstedt:2022zro,Lofgren:2023sep,Gould:2023ovu,Ekstedt:2024etx}, yet in order to describe the non-perturbative phenomena, one has to turn to lattice Monte Carlo simulations \cite{Farakos:1994xh,Kajantie:1995kf,Kajantie:1996qd, Laine:2000rm,Kainulainen:2019kyp,Niemi:2020hto,Gould:2021dzl,Gould:2022ran}.%
\footnote{For computation of the bubble nucleation rate utilising a thermal EFT, see e.g.~\cite{Gould:2021ccf, Ekstedt:2021kyx,Ekstedt:2022ceo,Ekstedt:2022tqk, Hirvonen:2021zej, Ekstedt:2023sqc} using perturbation theory, and \cite{Moore:2000jw,Moore:2001vf,Gould:2022ran,Gould:2024chm} using non-perturbative approaches.} 
Dimensionally reduced EFT for the xSM has previously been studied in \cite{Brauner:2016fla,Gould:2019qek,Schicho:2021gca,Niemi:2021qvp,Gould:2021oba,Tenkanen:2022tly,Niemi:2024axp,Niemi:2024vzw,Lewicki:2024xan}.
In essence, lattice simulations are crucial in order to answer two non-trivial questions: for a given BSM theory parameter space point 
\begin{itemize}
 
\item[] (i) is there a first-order phase transition? 

\item[] (ii) how accurately we can compute equilibrium thermodynamics and bubble dynamics, in perturbation theory?
\end{itemize}

Dismally, use of Monte Carlo lattice simulations is computationally expensive and labor-intensive, making them -- at first glance -- unsuitable for conducting broad surveys of phase transition thermodynamics in a parameter space of a complicated BSM model.  Recently, such lattice simulations have been limited to few parameter points -- \textit{benchmarks} of perturbation theory -- and have been performed 
in \cite{Laine:2012jy,Kainulainen:2019kyp, Niemi:2020hto,Niemi:2024axp}.
These benchmark studies have revealed that, fortunately, perturbation theory appears to predict first-order (scalar-driven) transitions correctly provided that a transition is strong enough, and perturbation theory is used at \textit{two-loop} order \cite{Gould:2023ovu}.

In this work at hand, we answer (i) by utilising the universality of the effective field theory at high-temperature that describes transition to the Higgs phase \cite{Kajantie:1995kf,Gould:2021dzl} (see Sec.~\ref{sec:3dEFT} for details). This strategy has been utilised before in e.g. \cite{Cline:1996cr,Andersen:2017ika, Niemi:2018asa,Gould:2019qek,Friedrich:2022cak}, and makes possible surveying phase structure in large parameter spaces in a wide range of models.  
In addition, when possible, we contrast our analysis to the latest lattice simulations of \cite{Niemi:2024axp} in the determination of regions of first-order phase transitions in the xSM. Once we have determined the character of a transition, and in particular confirmed the existence of first-order phase transitions, we employ perturbation theory within the thermal EFT. 

The same strategy based on thermal effective field theories has been employed previously 
in \cite{Gould:2019qek,Friedrich:2022cak} (see also \cite{Kierkla:2023von,Lewicki:2024xan}).    
In the present work at hand, we build upon the generic road map outlined in \cite{Friedrich:2022cak},
and further incorporate the following enhancements in computation of thermodynamics:
\begin{itemize}

    \item We include two-loop corrections to the thermal effective potential 
    within the EFT (in addition to two-loop thermal masses that are accounted when constructing the EFT).
    This allows for improved computation of the phase transition strength, and can substantially mitigate uncertainties to predictions of the GW signals \cite{Croon:2020cgk,Gould:2021oba}. 
 
    \item We calculate an estimate for the bubble wall velocity using local thermal equilibrium approximation \cite{Ai:2023see}. This estimate is obtained from the equilibrium pressure, which we compute at two-loop order, hence for the first time incorporating higher order thermal corrections to this quantity.

\end{itemize}
These major improvements allow us to significantly  develop the accuracy of the pipeline from xSM phenomenology to gravitational wave predictions compared to previous studies, e.g.\cite{Vaskonen:2016yiu,Beniwal:2017eik,Ellis:2018mja,Beniwal:2018hyi,Alves:2018jsw,Alanne:2019bsm,LISACosmologyWorkingGroup:2022jok}.%
\footnote{
One of the largest uncertainties in the pipeline still stems from the present low-order computation of the bubble nucleation rate at high temperatures \cite{Gould:2021oba}. Despite recent developments \cite{Ekstedt:2021kyx,Ekstedt:2023sqc}, higher order corrections to the bubble nucleation rate still remain as a major obstacle to overcome for uncorking the highest achievable, perturbative precision \cite{Ekstedt:2024etx}
} 
Our main results are shown in Figs.~\ref{fig:sintheta-a2} and~\ref{fig:mS-a2} and can be summarised as

\begin{itemize}

    \item The mass of the new scalar, $m_{h_2}$, portal coupling to the Higgs boson, $a_2$, and the mixing angle between the new scalar and the Higgs boson can exert significant influence on the phase structure diagram, as well as on thermodynamic and bubble dynamic quantities. Specifically, a first-order phase transition and viable nucleation necessitates a larger $a_2$ as $m_{h_2}$ becomes heavier and $\sin \theta$ decreases. The strength and duration of the EWPT 
show a notable sensitivity to the portal coupling $a_2$. This sensitivity propagates to the LISA signal-to-noise ratio, making detection possible only on narrow bands along free parameter space.

    \item We observe a strong correlation among the bubble wall velocity, nucleation temperature, and the phase transition strength. 
A larger nucleation temperature results in a smaller phase transition strength and bubble wall velocity. 
Notably, the bubble wall velocity spans the range of $[0.60, 0.9]$, indicating a hybrid profile solutions characterized by walls possessing both rarefaction and shock wave features.

    \item We find that the LISA sensitivity region favors a relatively large $\sin\theta$ value and specific ranges below $\sim 500$ GeV for the scalar mass. For instance, our results suggest $\sin\theta > 0.02$ (see Fig.~\ref{fig:sintheta-a2}), and the scalar mass spans from $230$ GeV to $395$ GeV and from $475$ GeV to $485$ GeV (see Fig.~\ref{fig:mS-a2}). Notably, we observe that the current measurements from the di-Higgs $b\bar{b}\tau^+\tau^-$ search at ATLAS can exclude a substantial portion of parameter space conducive to strong GW signals.
 
   \item A significant portion of the GW and collider probed regions can overlap, suggesting a simultaneous accountability for both signals if they are detected by future experiments. Conversely, depending heavily on the mixing angle, a null-result from collider experiments can potentially rule-out the xSM as an explanation for the GW detection.

\end{itemize}

Remainder of this article is organized as follows.
In Section~\ref{sec:model}, we define the model and discuss its collider phenomenology.
In Section~\ref{sec:3dEFT}, we discuss the high-temperature effective description for the model, its two-loop thermal effective potential and computation of thermal parameters for gravitational wave predictions.
We present our numerical results in Section~\ref{sec:numerical}, and discuss our findings further in Section~\ref{sec:conclusion}.
Constraints on the model are presented in Appendix~\ref{app:constraints}.
In Appendix~\ref{sec:matching}, we collect the matching relations between the thermal effective theory and the model at zero temperature.

\section{Model and collider phenomenology}
\label{sec:model}

We work on the real-singlet scalar extension of the SM~\cite{Profumo:2007wc}. 
The scalar potential at tree level is given by 
\begin{align}
\label{eq:4Dpot-tree}
V(\phi, S) = & \,\mu^2 \phi^{\dagger}\phi + \lambda (\phi^{\dagger}\phi)^2 + b_1 S + \frac12 b_2 S^2 \nonumber \\
& + \frac13 b_3 S^3 + \frac14 b_4 S^4  + \frac12 a_1 S \phi^{\dagger}\phi + \frac12 a_2 S^2 \phi^{\dagger}\phi, 
\end{align}
where $S$ is the real singlet scalar, transforming trivially under all SM gauge groups. 
The Higgs doublet $\phi$ can be parametrised as 
\begin{align}
\label{eq:doublet}
\phi = \begin{pmatrix} 
G^+ \\
\frac{1}{\sqrt{2}} \left( v_h + h + i G^0\right) 
\end{pmatrix}
\end{align}
where $G^{\pm}$, $G^0$ are the Goldstone bosons, $v_h \simeq 246$ GeV is the (gauge-fixed) electroweak vacuum expectation value (VEV) and $h$ is the real Higgs state.  
If $b_1 = b_3 = 0$ and $a_1 = 0$, the model has
a discrete $Z_2$ symmetry under $S \rightarrow -S$, unless the 
singlet acquires a VEV, leading to the spontaneous breaking of this symmetry. 
We note that, one can shift $S$ by a constant without changing the physical predictions of the theory \cite{OConnell:2006rsp, Profumo:2007wc, Espinosa:2011ax}.
Such shift is typically performed to either remove the tadpole term proportional to $b_1$, or to set singlet VEV $\langle S \rangle = 0$. 

In general, scalar states $h$ and $S$ can mix and form the mass eigenstates
$h_1$ and $h_2$ via the mixing matrix
\begin{align}
\begin{pmatrix} 
h_1 \\ h_2
\end{pmatrix}
= 
\begin{pmatrix} 
\cos\theta & \sin\theta \\
-\sin\theta & \cos\theta
\end{pmatrix}
\begin{pmatrix} 
h \\ S
\end{pmatrix}.
\end{align}
We identify the lighter mass eigenstate $h_1$ as the SM-like Higgs boson with the mass $m_{h_1} = 125.1$ GeV.
The Higgs data measurements at the LHC constraint the mixing angle $|\sin\theta| \lesssim 0.2$ \cite{CMS-PAS-HIG-19-005,ATLAS:2022vkf}. 
Further discussion on other experimental constraints on the new scalar mass and mixing angle can be found in Appendix~\ref{app:constraints}.

In this study, we concentrate on a scenario in which the $Z_2$ symmetry is explicitly broken, i.e. $b_1$, $b_3$ and $a_1$ are non-vanishing 
and we further fix $\langle S \rangle = 0$ at zero temperature. 
One then obtains the following relations for the tree-level potential parameters:  
\begin{align}
\label{eq:params-tree1}
\mu^2 =& - \frac12 \left( m^2_{h_1} \sin^2\theta + m^2_{h_2} \cos^2\theta \right),  \\
b_2 =& \, m^2_{h_1} \cos^2\theta + m^2_{h_2} \sin^2\theta -  \frac12 a_2 v_h^2 ,  \\
\lambda = & - \frac{\mu^2 }{v_h^2} ,  \\
a_1 =&\,  \frac{(m^2_{h_2} - m^2_{h_1}) \sin2\theta } {v_h}, \\
b_1 =& -\frac14 v_h^2  a_1. 
\end{align}
By this parametrisation, the free parameters in the model are the mass of the heavier (singlet-like) scalar boson $m_{h_2}$, the mixing angle $\theta$, the portal coupling $a_2$, and the singlet scalar self-interaction couplings $b_3$ and $b_4$. 
We note that for our following computation of thermodynamics in the next section, it is important to upgrade above tree-level relations to include one-loop quantum corrections at zero temperature \cite{Kajantie:1995dw}, and for this we follow \cite{Niemi:2021qvp}. 

The new scalar boson can decay into a pair of SM-like Higgs bosons if kinematically allowed ({\it i.e.}~$ m_{h_2} \geq 2 m_{h_1}$),  
and the produced pair of Higgs bosons can subsequently decay into SM particles. 
A study of the heavy resonance di-Higgs production in the context of xSM has been carried out in \cite{No:2013wsa, Chen:2014ask, Kotwal:2016tex, Huang:2017jws, Li:2019tfd, Zhang:2023jvh}. 
In our analysis we focus on the $b\bar{b}\gamma \gamma$ and $b\bar{b}\tau^+\tau^-$ final state channels. 
Probing these two channels have been carried out at the LHC Run 2 including ATLAS \cite{ATLAS:2021ifb, ATLAS:2018dpp, ATLAS-CONF-2021-030} and CMS \cite{CMS:2018tla, CMS:2017hea} searches and have provided the most stringent bounds on the extra scalar boson mass below $\sim 500$ GeV. In particular, 
The ${b\bar{b} \gamma \gamma}$ search is the most sensitive at low resonance mass ($< 320$ GeV) while the $b\bar{b}\tau^+\tau^-$ is more sensitive for a higher resonance mass. 

The cross section for the process $pp \to h_2 \to h_1 h_1 \to b\bar{b}\gamma \gamma$ ($ b\bar{b}\tau^+\tau^-$) in a narrow width approximation is given by 
\bea
\sigma_{b\bar{b} \gamma \gamma} &=& \sigma_{pp\to h_2}\times {\rm BR}(h_2 \to h_1 h_1) \times {\rm BR}(h_1 \to  b\bar{b} ) \times {\rm BR}(h_1 \to \gamma \gamma),  \\ 
\sigma_{b\bar{b} \tau^+\tau^-} &=& \sigma_{pp\to h_2}\times {\rm BR}(h_2 \to h_1 h_1) \times {\rm BR}(h_1 \to  b\bar{b} ) \times {\rm BR}(h_1 \to \tau^+\tau^-). 
\eea
Here the branching ratios of the di-Higgs are given as ${\rm BR} (h_1 \to b {\bar b}) \times {\rm BR}(h_1 \to \gamma \gamma) \simeq 0.13\%$ and ${\rm BR} (h_1 \to b {\bar b}) \times {\rm BR}(h_1 \to \tau^+ \tau^-) \simeq 3.67\%$~\cite{LHCHiggsCrossSectionWorkingGroup:2016ypw}. 
The production cross section of $h_2$ is 
$\sigma_{pp\to h_2} = \sin^2\theta \times \sigma^{\rm SM}(pp\to H)|_{m_H = m_{h_2}}$, where the SM cross section $\sigma^{\rm SM}(pp\to H)$ can be obtained from \cite{Cepeda:2019klc}. 
The branching ratio of $h_2 \to h_1 h_1$ can be given as 
\begin{equation}
{\rm BR}(h_2 \to h_1 h_1)=\frac{\Gamma_{h_2 \to h_1 h_1}}{\Gamma_{h_2 \to h_1 h_1}+\sin^2\theta ~\Gamma^{\rm SM}(m_{h_2})}.
\end{equation}
Here $\Gamma^{\rm SM}(m_{h_2})$ is the SM-like Higgs boson decay width evaluated at $m_{h_2}$.
The partial width of $h_2 \to h_1 h_1$ is given by 
\begin{equation}
\Gamma_{h_2 \to h_1 h_1}=\frac{\lambda_{211}^2 \sqrt{1-\frac{4m_{h_1}^2}{m_{h_2}^2}}}{32\pi m_{h_2}},
\end{equation}
where the cubic coupling 
\be
\lambda_{211} \equiv 2 s_\theta^2 c_\theta b_3 + \frac{a_1}{2} c_\theta (c_\theta^2 - 2 s_\theta^2) + (2 c_\theta^2 - s_\theta^2) s_\theta v_h a_2 - 6 \lambda s_\theta c_\theta^2 v_h, 
\ee
with shorthand notations $s_\theta \equiv \sin\theta$ and $c_\theta \equiv \cos\theta$. 

For the $b\bar{b}\tau^+\tau^-$ final state channel, we recast recent results from ATLAS Run 2 \cite{ATLAS-CONF-2021-030} with the luminosity of $139\,{\rm fb}^{-1}$. To obtain the HL-LHC sensitivity, we rescale the current ATLAS limits by a factor of $\sqrt{139\,{\rm fb}^{-1}/(1.18\times {\cal L})}$ where 1.18 is a factor accounting for increasing the center-of-mass energy $\sqrt{s} = 13$ TeV to 14 TeV and the luminosity at HL-LHC is $ {\cal L} = 3000\, {\rm fb}^{-1}$. 

For the $b\bar{b}\gamma\gamma$ final state channel,
the number of signal events at the HL-LHC can be obtained by 
\be
n^{b\bar{b} \gamma \gamma}_s = {\cal L}\times \sigma_{b\bar{b} \gamma \gamma} \times {\cal A}\times \epsilon. 
\ee
where ${\cal A}\times \epsilon$ is the acceptance times efficiency for the detection.
We follow an analysis strategy in \cite{ATLAS:2018dpp} and focus on the \texttt{2-tag category} signal region where the events consist exactly two $b$-jets satisfying the requirement for the $70\%$ efficient working point.%
\footnote{
For the resonance searches, this signal region yields a higher detection efficiency compared to \texttt{1-tag category} \cite{ATLAS:2018dpp}. } 
In this signal region, the acceptance times efficiency ranges from $6\%$ to $15.4\%$ for the resonance mass ranges from $260$ GeV to $1000$ GeV \cite{ATLAS:2018dpp}. 
The number of background event for the ${b\bar{b} \gamma \gamma}$ final state, denoted as $n_{b}^{{b\bar{b} \gamma \gamma}}$, at the HL-LHC is obtained by rescaling the number of background event taken from \cite{ATLAS:2018dpp} to the luminosity at the HL-LHC.
The significance is then given by
\be
{\cal Z}_{b\bar{b} \gamma \gamma} = \sqrt{2 (n^{b\bar{b} \gamma \gamma}_s + n^{b\bar{b} \gamma \gamma}_b) \log\left( 1+ \frac{n^{b\bar{b} \gamma \gamma}_s} {n^{b\bar{b} \gamma \gamma}_b}\right) - 2 n^{b\bar{b} \gamma \gamma}_s}.
\ee

\section{Thermodynamics and gravitational waves}
\label{sec:3dEFT}

At high temperatures ($T$), non-perturbative physics arise due to high occupancy of bosonic modes: consider a loop expansion with a generic weak coupling $g^2 \ll 1$. When $T \gg E$, the effective expansion parameter is not just $g^2$ but rather $g^2 n_B(E,T) \geq g^2 \frac{T}{m}$, since each loop order comes with the associated Bose-Einstein distribution $n_B$, where energy (mass) of each mode is denoted by $E$ ($m$). Therefore, low energy -- or long-distance/infrared -- bosonic excitations with $m \sim g^2 T$ become Bose-enhanced and strongly coupled at high temperatures, i.e. the effective expansion parameter $g^2 n_B \sim \mathcal{O}(1)$ \cite{Linde:1980ts,Gross:1980br}. The infrared (IR) regime $m \sim g^2 T$ is beyond the reach of perturbation theory, and this poses a problem for   
the high-temperature ``symmetric phase'': in this phase, non-abelian SU(2) gauge fields experience confinement, which leads to massive vector-boson bound states -- with so-called ``magnetic mass'' at $\mathcal{O}(g^2 T)$ \cite{Shaposhnikov:1993jh,Irback:1986dc,Farakos:1986tx}. 
In the ``broken phase'', on the other hand, due to the Higgs mechanism (i.e. condensation of a scalar field) masses for vector-boson excitations become large compared to the non-perturbative magnetic mass.
Hence, the perturbative description of the broken phase becomes possible, due to the infrared cutoff provided by the Higgs condensation%
\footnote{
Along the lines of recent \cite{Niemi:2024axp}, we use terms symmetric and broken phases, while keeping in mind that there is no gauge invariant order parameter to distinguish the phases \cite{Elitzur:1975im}. 
}.
To determine the critical temperature, the free-energy of both phases, however, is required.

In principle, non-perturbative simulations within a dimensionally reduced EFT are straightforward \cite{Farakos:1994xh}, compared to direct simulations of a parent theory at high temperatures \cite{Csikor:1998eu}: in the EFT there are no issues with simulating chiral fermions, as the EFT in three-dimensions is purely bosonic, with fermions integrated out perturbatively at high temperatures. In addition, EFT simulations have to resolve fewer length scales due to reduced dimensions, and relations between lattice and continuum theories are known exactly due to super-renormalisability of the EFT at three-dimensions \cite{Laine:1995np,Laine:1997dy}.

On the other hand, purely perturbative studies are significantly less computationally demanding. As shown in recent study \cite{Gould:2023ovu}, the characteristic mass scale for a strong phase transition lies above the non-perturbative magnetic mass scale, meaning it can be accurately described by perturbation theory, provided the expansion is carried out to sufficiently high orders. However, the convergence of the perturbative expansion, particularly for quantities like the free energy at high temperatures, is much slower than at zero temperature. In practice, this requires two-loop computations to achieve reliable results \cite{Ekstedt:2024etx}.

In our approach, we combine the two approaches, by repurposing results of previous lattice simulations to confirm the existence of a first-order phase transition, and once scrutinized, resorting to perturbative computations thereafter.
Phase diagram of the SU(2) + Higgs EFT is known at non-perturbative level from simulations of \cite{Kajantie:1995kf,Gould:2021dzl}, and any generic BSM theory with massive enough scalars and feeble portal interaction to the Higgs at high temperatures maps into this thermal EFT, describing a smooth crossover in analog to minimal Standard Model. In order to make the transition first-order, portal interactions with Higgs and new BSM scalars need to be large enough to reduce the effective, thermal self-interaction coupling of the Higgs.%
\footnote{This strategy allows to find a boundary in model parameter space between regions of smooth crossover and first-order phase transitions, for electroweak phase transitions proceeding in a single step from symmetric to broken electroweak phase. However, it also gives general information about locations of multi-step phase transitions, as these require lighter BSM scalars that cannot be integrated out, and hence these regions cannot overlap with those regions where mapping into single Higgs EFT describes a crossover.}
The effects of BSM physics are fully captured in the EFT matching relations, and are purely perturbative. 

\subsection{Thermal effective field theory}

We adopt the methodology outlined in \cite{Schicho:2021gca,Niemi:2021qvp} to formulate a thermal effective field theory and consequently thermal effective potential as well as the effective action for the xSM at high temperatures. 
In short, we utilise the scale hierarchies present at high temperatures and integrate out the non-zero Matsubara modes from ``hard'' scales $\sim \pi T$, to construct a three-dimensional EFT at ``soft'' scale $\sim g T$. 
Notably, this dimensional reduction procedure includes all essential thermal resummations and effectively sums up a subset of higher-order corrections through the renormalization group. In particular, we are able include two-loop thermal masses and other perturbative effects at same order in high-temperature expansion, that have been reported to be crucially important for quantitatively reliable analysis \cite{Laine:2012jy,Kainulainen:2019kyp,Croon:2020cgk,Gould:2021oba}.

Additionally, we integrate out the temporal gauge field components, that are Lorentz scalars with characteristic Debye mass scale of $g T$, thereby leaving us with an EFT at final, ``softer'' scale%
\footnote{Often in the literature, the scale of the final EFT is referred as the ``ultrasoft'' scale $\sim g^2 T$ \cite{Kajantie:1995dw}, which is the non-perturbative scale of the magnetic mass of non-abelian gauge bosons at high temperatures. However, as argued in \cite{Gould:2023ovu}, scalars that drive the first-order phase transition live at scale $\mathcal{M}$, such that $g T \gg \mathcal{M} \gg g^2 T$. In a context of radiatively generated barriers, it is natural to assign this mass scale through the geometric mean $\mathcal{M} \sim g^{\frac{3}{2}}T$, dubbed as the ``supersoft'' scale. In our analysis, we refrain from attaching any formal power counting for the ``softer'' scale $\mathcal{M}$. }
with Higgs, singlet and spatial gauge fields. 
We relegate precise definitions of the EFT parameters to Appendix~\ref{sec:matching}, where we also -- for the reader's benefit -- collect all EFT matching relations. These relations have originally been computed in \cite{Schicho:2021gca,Niemi:2021qvp}, and can also be obtained with the public {\tt DRalgo} code \cite{Ekstedt:2022bff}, which can be used for other, generic models as well.

\subsection{Thermal effective potential}

Within the EFT, the tree-level effective potential reads 
\begin{align}
\label{eq:Veff-tree}
V_{0} &=   \frac12 \bar{\mu}^2_{3} \bar{v}^2 + \frac14 \bar{\lambda}_3 \bar{v}^4  + \frac14 \bar{a}_{1,3} \bar{v}^2 \bar{s} + \frac14 \bar{a}_{2,3} \bar{v}^2 \bar{s}^2  + \bar{b}_{1,3} \bar{s} + \frac12 \bar{b}_{2,3} \bar{s}^2 + \frac13 \bar{b}_{3,3} \bar{s}^3 + \frac14 \bar{b}_{4,3} \bar{s}^4 
\end{align}
where 
parameters denoted by bar are the effective, temperature-dependent parameters, and the Higgs and singlet background fields are denoted as $\bar{v}$ and $\bar{s}$, respectively.   
The one-loop correction to the effective potential reads
\bea
V_1 &=& 2(d - 1) J_3(\bar{m}_W) + (d - 1)J_3(\bar{m}_Z) +  3 J_3(\bar{m}_G) \nonumber \\ 
&+&  J_3(\bar{m}_{h_1}) + J_3(\bar{m}_{h_2}),
\eea
where dimension of space $d = 3 - 2\epsilon$ in dimensional regularisation and the one-loop integral reads
\be
J_3(m) = \frac12 \Big( \frac{e^\gamma \Lambda^2}{4\pi} \Big)^\epsilon \int \frac{d^d p}{(2\pi)^d} \ln (p^2 + m^2) = -\frac{(m^2)^{\frac{3}{2}}}{12\pi},
\ee
in the minimal subtraction scheme, with Euler-Mascheroni constant $\gamma$ and the renormalisation scale $\Lambda$.%
\footnote{In three-dimensions, this one-loop integral is UV finite and hence does not depend on the renormalisation scheme: UV divergences appear at two-loop order.}
It is this one-loop contribution, which -- if truncated to leading order in matching relations -- exactly matches the frequently appearing ring- or daisy-resummation of \cite{Arnold:1992rz}, see e.g.~Appendix A of \cite{Niemi:2024vzw}. 

Mass (squared) eigenvalues of the gauge and Goldstone bosons are given by 
\begin{align}
\bar{m}_W^2 &= \frac14 \bar{g}^2_3 \bar{v}^2 , \quad\quad \bar{m}_Z^2 = \frac14 \left(\bar{g}^2_3 + (\bar{g}'_3)^2 \right) \bar{v}^2 \\
\bar{m}_G^2 &= \bar{\mu}^2_{3} + \bar{\lambda}_3 \bar{v}^2 + \frac12 \bar{a}_{1,3} \bar{s} + \frac12 \bar{a}_{2,3} \bar{s}^2. 
\end{align}
Mass eigenvalues of the scalar bosons $\bar{m}^2_{h_1}$ and $\bar{m}^2_{h_2}$ 
are obtained by diagonalizing the background field dependent mass matrix
\bea
\label{eq:scalarmatrix}
M &\equiv& 
\begin{pmatrix}
M_{11} & M_{12} \\
M_{12} & M_{22}
\end{pmatrix}
 = 
\begin{pmatrix}
\bar{\mu}^2_3 + 3 \bar{\lambda}_3 \bar{v}^2 + \frac12 \bar{a}_{1,3} \bar{s} + \frac12 \bar{a}_{2,3} \bar{s}^2 
&  \left(\frac12 \bar{a}_{1,3} + \bar{a}_{2,3} \bar{s} \right)\bar{v} \\
 \left(\frac12 \bar{a}_{1,3} + \bar{a}_{2,3} \bar{s} \right)\bar{v} &
 \frac12 \bar{a}_{2,3} \bar{v}^2 + \bar{b}_{2,3} + 2 \bar{b}_{3,3} \bar{s} + 3 \bar{b}_{4,3} \bar{s}^2 
\end{pmatrix} \; .\nonumber 
\\
\eea
as $R^T \cdot M \cdot R = {\rm diag}(\bar{m}_{h_1}^2, \bar{m}_{h_2}^2)$, with the orthogonal rotation matrix
\bea
R &=&
\begin{pmatrix}
\cos {\bar{\theta} } & \sin {\bar{\theta}} \\
- \sin {\bar{\theta} } & \cos {\bar{\theta} } 
\end{pmatrix} \; ,
\eea
with the mixing angle
\be
\sin(2 {\bar{\theta} }) = \frac{2 M_{12}} {\sqrt{\left(M_{11} - M_{22} \right)^2 + 4 M_{12}^2 }}.
\ee
This results
\be
{\bar{m} }^2_{h_1,h_2} = \frac12 \left( M_{11} + M_{22} \mp \sqrt{\left(M_{11} - M_{22} \right)^2 + 4 M_{12}^2 } \right).
\ee
To improve the precision in determining the thermodynamic quantities, we include two-loop corrections to the effective potential within the EFT. The expression for this two-loop potential, $V_2$, can be read from the appendix B of \cite{Niemi:2021qvp}; for brevity, we refrain from reproducing it here. 

To guarantee the gauge invariance of our further calculations, 
we employ the so-called ``$\hbar$-expansion'' and follow \cite{Laine:1994zq,Niemi:2020hto, Croon:2020cgk}
by expanding the effective potential 
order-by-order in the loop-counting parameter $\hbar$. 
To quadratic order in $\hbar$, the expansion of the potential and the background field at its minima read formally
\bea
\label{eq:Veff-expand}
V_{\rm eff}^{\hbar} &=& V_0 + \hbar V_1 + \hbar^2 V_2, \quad \\
 \bar{v}_\text{min} &=& \bar{v}_0 + \hbar \bar{v}_1 + \hbar^2 \bar{v}_2, \quad \\
 \bar{s}_\text{min} &=& \bar{s}_0 + \hbar \bar{s}_1 + \hbar^2 \bar{s}_2,
\eea
where $\pdv{V_0}{\bar{v}}|_{\bar{v} = \bar{v}_0} = 0$ and $\pdv{V_0}{\bar{s}}|_{\bar{s} = \bar{s}_0} = 0$, {\it i.e.} the leading order solutions $\bar{v}_0$ and $\bar{s}_0$ extremize the tree-level potential $V_0$. In general, for the extrema $(\bar{v}_0, \bar{s}_0)$ there are 9 different solutions, some of them being physically equivalent.
By evaluating the effective potential at its minima and expanding, we obtain%
\footnote{
The expansion of Eq.~\eqref{eq:Vhbar} relies on the fact that minima $(\bar{v}_0, \bar{s}_0)$ are separated by a potential barrier, which is already present at tree-level in $V_0$, due to the non-zero cubic portal $\bar{a}_{1,3}$. For alternative, effective field theory expansions we refer discussion in \cite{Gould:2023ovu}. 
}
\bea
\label{eq:Vhbar}
V_{\rm eff}^{\hbar}(\bar{v}_\text{min}, \bar{s}_\text{min}) &=& V_0(\bar{v}_0, \bar{s}_0) + \hbar V_1(\bar{v}_0, \bar{s}_0)  \nonumber \\ 
&& \hspace{-2.5cm} +\, \hbar^2 \left[V_2(\bar{v}_0, \bar{s}_0) - \frac12 \bar{v}_1^2 \pdv[2]{V_0}{\bar{v}} - \frac12 \bar{s}_1^2 \pdv[2]{V_0}{\bar{s}} - \bar{v}_1 \bar{s}_1 \pdv{V_0}{\bar{v}}{\bar{s}} \right] + \,\mathcal{O}(\hbar^3) ,
\eea
where ${\cal O}(\hbar)$  corrections for the minima are given as
\begin{align}
\label{eq:v1s1}
\bar{v}_1 =& \left[\left( \pdv{V_0}{\bar{v}}{\bar{s}} \right)^2 - \left(\pdv[2]{V_0}{\bar{v}}\right) \left(\pdv[2]{V_0}{\bar{s}} \right) \right]^{-1} \left[ \left( \pdv[2]{V_0}{\bar{s}} \right)\left( \pdv{V_1}{\bar{v}} \right) - \left( \pdv{V_0}{\bar{v}}{\bar{s}} \right)\left( \pdv{V_1}{\bar{s}} \right) \right], \\
\bar{s}_1 =& \left[\left( \pdv{V_0}{\bar{v}}{\bar{s}} \right)^2 - \left(\pdv[2]{V_0}{\bar{v}}\right) \left(\pdv[2]{V_0}{\bar{s}} \right) \right]^{-1} \left[ \left( \pdv[2]{V_0}{\bar{v}} \right)\left( \pdv{V_1}{\bar{s}} \right) - \left( \pdv{V_0}{\bar{v}}{\bar{s}} \right)\left( \pdv{V_1}{\bar{v}} \right) \right],
\end{align}
and all derivatives are evaluated at the tree-level minima $(\bar{v}_0, \bar{s}_0)$. The expansion of Eq.~\eqref{eq:Vhbar} satisfies Nielsen-Fukuda-Kugo identities~\cite{Nielsen:1975fs, Fukuda:1975di}, and is therefore gauge-invariant order by order.

To study the phase structure in the model, we determine the evolution of $V_{\rm eff}^{\hbar}$ as a function of temperature in different phases \cite{Patel:2011th,Schicho:2022wty}. The critical temperature $T_c$ can be determined from the condition that $V_{\rm eff}^{\hbar}$ in any two minima are degenerate. 
We focus on \textit{two-step} transitions for which the symmetry breaking pattern is schematically given as 
\be
\label{eq:PTpattern}
(\bar{v} = 0, \bar{s} \simeq 0) \to (\bar{v} = 0, \bar{s} \neq 0) \to (\bar{v} \neq 0, \bar{s} \simeq 0). 
\ee
Here both first and second steps of a transition can give rise to a first-order phase transition where the barrier between the minima is mainly generated from tree-level effects. In our analysis below, we will only concentrate on the second step of the transition, as these transitions are both stronger, and slower, which leads to more promising prospects for gravitational wave production.

\subsection{Effective action and bubble nucleation rate}

The first-order EWPTs proceed through the nucleation of bubbles of a stable phase, which grow until they eventually supplant the pre-existing metastable phase. 
The collisions of the bubbles, and even more so the subsequent fluid dynamics in the plasma, produce the shear stresses that source gravitational waves. 
In the following, we review the expressions of required quantities for the prediction of the GW spectrum from the first-order EWPT. Analogous treatments of the bubble nucleation rate using thermal EFT approach can be found in \cite{Croon:2020cgk,Gould:2021oba, Gould:2021ccf,Friedrich:2022cak,Ekstedt:2024etx}. 

The thermal bubble nucleation rate at leading approximation reads 
\begin{align}
\Gamma = A(T) e^{- S_{\rm eff}^{\rm LO}(\Phi_{3})},
\end{align} 
where the ``classical'' contribution%
\footnote{
The exponential with the leading order action is the \textit{classical} rate within the EFT, that is constructed by integrating out heavy thermal fluctuations \cite{Ekstedt:2022tqk}.  
}
is described by the exponential with $S_{\rm eff}^{\rm LO}$, i.e. the leading order action, and dominant over the prefactor $A(T) \sim T^4$, that composes of (one-loop) quantum fluctuation determinant, of both scalar and gauge fields \cite{Ekstedt:2021kyx}. 
The leading order action reads
\begin{align}
\label{eq:action}
S_{\rm{eff}}^{\rm LO}(\Phi_{3}) &= \int_{0}^{\infty} dr r^2 \frac{1}{2} \left( \pdv{\Phi_3}{r} \right)^2 + V_0(\Phi_3, T).
\end{align} 
Here $\Phi_{3} = \{\bar{v}, \bar{s}\}$ denotes a collection of the background fields, that minimizes the action $\mathcal S_{\rm{eff}}^{\rm LO}$ and are determined by solving the following equation of motion, 
\begin{equation}
\label{eq:bounceeq}
\frac{{\rm d}^2 \Phi_{3}}{{\rm d} r^2}+\frac{2}{r} \frac{{\rm d} \Phi_{3} }{{\rm d} r}=\frac{{\rm d} V_{0}(\Phi_{3}, T)}{{\rm d} r} \,, 
\end{equation}
with the boundary conditions
\begin{equation}
\lim_{r \rightarrow \infty} \Phi_3(r) = 0,\left.\quad \frac{{\rm d} \Phi_3}{{\rm d} r}\right|_{r=0}=0 \,. 
\end{equation}
We utilize the Mathematica package {\tt FindBounce}~\cite{Guada:2020xnz} to numerically solve the bounce equation in (\ref{eq:bounceeq}) and then evaluate the action in~(\ref{eq:action}).

We emphasize, that due to expansion of Eq.~\eqref{eq:Vhbar}, only the tree-level potential $V_0$ appears in the leading effective action in (\ref{eq:action}).
Hence, our computation is automatically gauge invariant \cite{Croon:2020cgk}.
Using $V_0$ is possible, as the barrier separating the minima required by the bounce to exist, is present already at tree-level. 
This is in contrast to refs.~\cite{Hirvonen:2021zej,Lofgren:2021ogg,Ekstedt:2022ceo,Gould:2022ran,Kierkla:2023von} that consider radiatively generated barriers. Such barriers are generated by gauge boson fluctuations, which are heavier than the scalar undergoing the transition. This allows to handle the gauge boson fluctuations in derivative expansion, and consequently include their effects also at higher, two-loop order, which leads to a first possible renormalisation group (RG) improvement \cite{Hirvonen:2021zej,Ekstedt:2022ceo,Kierkla:2023von}, and such treatment has been shown to be gauge invariant in \cite{Hirvonen:2021zej,Lofgren:2021ogg}.
In our present treatment, gauge boson fluctuations are incorporated in the prefactor $A(T)$ \cite{Baacke:1999sc}, which -- at the present stage -- we do not compute.%
\footnote{
In principle, one could use {\tt BubbleDet} \cite{Ekstedt:2023sqc} to compute the prefactor, but currently multifield potentials are not yet supported by this code.  
} 
Indeed, in the expansion of Eq.~\eqref{eq:Vhbar}, the gauge field contributions appear only at next-to-leading order, but should they be comparable to tree-level terms responsible for the barrier, it is tempting to speculate on a possibility wherein these gauge field contributions are resumed to the leading order action, in analogy to treatment in \cite{Gould:2023ovu}. Such a treatment could facilitate RG improvement, mitigating the major bottleneck required for high accuracy determination of GW spectra, as reported in \cite{Gould:2021oba}. We leave such considerations for future work.  

Moving on, the inverse duration of the phase transition can be determined from
\begin{equation}
\frac{\beta}{H_*} = \left. -T \frac{d}{dT} \ln \Gamma \right|_{T = T_*}  \approx \left. T \frac{d S^{\rm LO}_{\rm eff}}{dT}\right|_{T = T_*},  
\end{equation} 
where $H_*$ represents the Hubble rate at temperature $T_*$,
and we omit (subleading) contribution of the prefactor $-T \frac{d}{dT} \ln A(T)$.
Temperature $T_*$ -- the temperature for GW production \cite{Caprini:2019egz} -- is identified with
the percolation temperature $T_p$, at which the condition $h(t_p) = 1/e$ is met, with $h(t_p)$ representing the fraction of space at the percolation time $t_p$ \cite{Athron:2023rfq}.
Solving the percolation condition approximately yields a condition \cite{Enqvist:1991xw, Caprini:2019egz} 
\begin{align}
S^{\rm LO}_{\rm eff}(T_*)  \simeq  131 + \log(\frac{A}{T_*^4}) - 4 \log(\frac{T_*}{100 {\rm GeV}}) - 4 \log(\frac{\beta/H_*}{100}) + 3 \log(v_w), 
\end{align}
where $v_w$ is the bubble wall velocity, from which $T_*$ can be solved.

Next, the phase transition strength, denoted by $\alpha$, quantifies the difference between the trace anomaly in the false vacuum $\theta_f$ and that in the true vacuum $\theta_t$, weighted by the enthalpy density $\omega_f$ in the false vacuum at $T_*$, i.e.
\begin{align}
\alpha = \left.\frac{\left[\theta_f(T) - \theta_t(T)\right]}{3 \omega_f (T)}\right|_{T = T_*}.  
\end{align}
Here, the trace anomaly and enthalpy density can be derived from the pressure $p(T)$ using the standard relations
\bea
\omega(T) &=& T \frac{\partial{p} }{ \partial{T}}, \\
\label{eq:thetaT}
\theta(T) &=& \rho(T) - 3p, 
\eea
where $\rho(T) = T \frac{\partial{p}}{\partial{T}} - p$ describes the energy density. We note that $\theta(T)$ in (\ref{eq:thetaT}) is taken in the relativistic plasma limit and in practice receives further corrections if the speed of sound differs from $c_s^2 = 1/3$ \cite{Giese:2020rtr, Tenkanen:2022tly}. 
At leading order the pressure is given by
\be
p(T) = \frac{\pi^2}{90} g_{*}T^4 - T V_{\rm eff}^{\hbar},
\ee
where $g_{*} = 106.75 + 1$ denotes the number of relativistic degrees of freedom, with an additional one accounted for the singlet degree of freedom.
Effective potential at higher orders captures the higher order corrections to the pressure from within the thermal EFT, while higher order corrections to the symmetric phase $T^4$-part can be accounted as in \cite{Gynther:2005dj,Gynther:2005av,Tenkanen:2022tly}.

\subsection{Bubble wall velocity in local thermal equilibrium approximation}

The last crucial ingredient for predictions of the GW spectrum is the terminal bubble wall velocity $v_w$. This quantity describes the speed of the phase interface after nucleation in the plasma's rest frame, distant from the wall. Our previous study \cite{Friedrich:2022cak} has shown a significant impact of the bubble wall velocity on the GW signal-to-noise ratio (SNR) for LISA detector, particularly noting that this peaks at $v_w \sim 0.63$. 
Determining the wall velocity poses a challenge due to the necessity of performing out-of-equilibrium calculations \cite{Moore:1995si, Konstandin:2014zta, Kozaczuk:2015owa, Dorsch:2018pat, Hoche:2020ysm,Laurent:2022jrs}. 

However, the out-of-equilibrium effects can be subdominant, especially in the context of the xSM \cite{Laurent:2022jrs}. Therefore, we can estimate the bubble wall velocity in straightforward manner by assuming a local thermal equilibrium (LTE) scenario and applying the conservation of entropy \cite{Ai:2021kak,Ai:2023see}. 
Recent hydrodynamic simulation results for the bubble-wall velocity under LTE in \cite{Krajewski:2024gma} have demonstrated a good agreement with those obtained with the analytical method in \cite{Ai:2023see}. 
Hence, we adopt the approximate form of the bubble wall velocity in LTE, by following \cite{Ai:2023see}
\be
\label{eq:vw}
v_w^{\rm LTE} = \left(\left| \frac{3\alpha + \Psi - 1}{2(2 - 3\Psi + \Psi^3)}\right|^{c/2} + \left| v_{\rm CJ} \left(1 - a \frac{(1-\Psi)^b}{\alpha} \right)\right|^c \right)^{1/c}. 
\ee
Here numerical fit constants have values a = 0.2233, b = 1.074, c = -3.433, and $\Psi = \omega_t/\omega_f$ is the ratio of enthalpies and the Chapman–Jouguet velocity $v_{\rm CJ}$ is given by 
\be
v_{\rm CJ} = \frac{1}{\sqrt{3}} \frac{1 + \sqrt{3\alpha^2 + 2\alpha}}{1+\alpha}. 
\ee
We note that the bubble wall velocity estimated from (\ref{eq:vw}) should be smaller than the Chapman–Jouguet velocity, i.e. $v_w^{\rm LTE} < v_{CJ}$ \cite{Ai:2023see}. 
We emphasize, that since we are able to include higher order corrections consistently for $\alpha$ and the pressure, as consequence we can compute $v_w^{\rm LTE}$ at higher orders as well. For many points, we find that two-loop corrections increase $\alpha$, and this results in increased value for $v_w^{\rm LTE}$.

\subsection{Signal-to-noise ratio for LISA}

Finally, given the thermal parameters $(T_*,\alpha_*,\beta/H_*, v_w)$ described above, we use the {\tt PTPlot} package \cite{Hindmarsh:2017gnf, Caprini:2019egz} to compute the GW spectrum. To assess the detectability of the signals, one can define the SNR \cite{Caprini:2019egz} as follows
\begin{equation}
\mathrm{SNR}=\sqrt{\mathcal{T} \int_{f_{\min }}^{f_{\max }} \mathrm{d} f\left[\frac{h^2 \Omega_{\mathrm{GW}}(f)}{h^2 \Omega_{\mathrm{exp}}(f)}\right]^2},
\end{equation}
where $\mathcal{T}$ represents the duration of the observation period in years, $h^2 \Omega_{\mathrm{GW}}(f)$ denotes the spectrum of the fraction of GW energy from the first-order phase transition, and $h^2\Omega_{\mathrm{exp}}(f)$ corresponds to the sensitivity of the experimental setup. 
For relativistic hydrodynamic simulations of GW production from first order phase transitions, see  \cite{Hindmarsh:2013xza,Hindmarsh:2015qta,Hindmarsh:2017gnf,Cutting:2018tjt,Cutting:2019zws,Cutting:2020nla,Dahl:2021wyk}.

\section{Numerical analysis}
\label{sec:numerical}

\begin{figure}[t]
   \centering
   \includegraphics[width=0.55\textwidth]{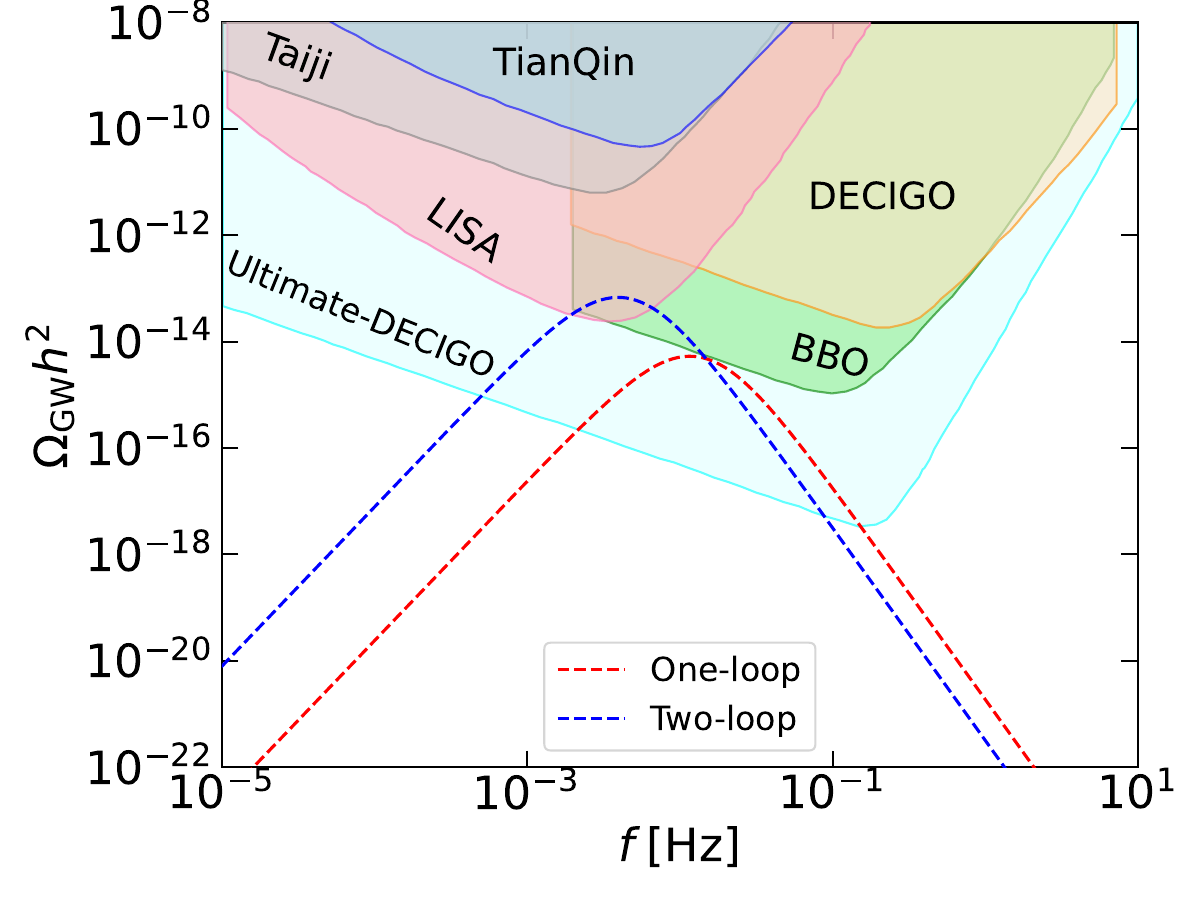}
   \caption{
    \label{fig:spectrum} The GW spectrum generated for the benchmark point with $m_{h_2} = 350$ GeV, $b_3 = 40$ GeV, $b_4 = 0.3$, $a_2 = 3.0$, and $\sin\theta = 0.1$, including one-loop (dashed red line) and two-loop (dashed blue line) corrections to the effective scalar potential. The shaded regions indicate the experimental sensitivities of various GW detectors, including Taiji~\cite{Gong:2014mca, Hu:2017mde, Guo:2018npi}, TianQin~\cite{TianQin:2015yph, Hu:2017yoc}, (Ultimate-)DECIGO~\cite{Kudoh:2005as,Kawamura:2011zz, Musha:2017usi}, BBO~\cite{Crowder:2005nr, Yagi:2011wg}, and LISA~\cite{Audley:2017drz,Cornish:2018dyw}.}
\end{figure}

\subsection{The effects of higher order corrections: a benchmark}

For starters, we examine the effects of two-loop corrections on the thermal parameters and graviational wave SNR as compared with the one-loop corrections only. For this, we consider a benchmark point with input parameters fixed as: $m_{h_2} = 350$ GeV, $b_3 = 40$ GeV, $b_4 = 0.3$, $a_2 = 3.0$, and $\sin\theta = 0.1$. At one-loop order, we find
\be
T_{*} = 82.12 \, \text{GeV}, \quad 
\alpha = 0.065, \quad 
\frac{\beta}{H_*} = 1102.57, \quad
v_w^{\text{LTE}} = 0.70, \quad
\text{SNR}_{\text{LISA}} = 0.18.
\ee
Including two-loop corrections results:
\be
T_{*} = 64.75 \, \text{GeV}, \quad 
\alpha = 0.128, \quad 
\frac{\beta}{H_*} = 528.4, \quad
v_w^{\text{LTE}} = 0.78, \quad
\text{SNR}_{\text{LISA}} = 9.3.
\ee
We observe that for this benchmark point, the LISA SNR is significantly enhanced -- by about two orders of magnitude -- when incorporating two-loop thermal corrections. 

In Fig.~\ref{fig:spectrum}, for the above benchmark point, we plot the GW spectrum using one-loop (dashed red line) and two-loop (dashed blue line) level computations, together with the experimental sensitivities for various future detectors.
The spectrum with two-loop thermal corrections has a higher peak and a lower frequency, potentially making it accessible to Ultimate-DECIGO~\cite{Kudoh:2005as,Kawamura:2011zz, Musha:2017usi}, BBO~\cite{Crowder:2005nr, Yagi:2011wg}, and LISA~\cite{Audley:2017drz,Cornish:2018dyw} detectors.
In contrast, in one-loop computation the spectrum peaks at a significantly lower and a slightly higher frequency, falling only within the potential detection range of the Ultimate-DECIGO detector. 

This particular benchmark study illustrates the importance of the two-loop thermal corrections, and demonstrates that perturbation theory at higher orders can lead to significantly stronger signals for GW experiments (see also \cite{Croon:2020cgk,Gould:2021oba,Gould:2023jbz,Lewicki:2024xan}). We observe the same trend for many other parameter points as well, but emphasize, that this trend is by no means general: a recent, exhaustive study \cite{Niemi:2024vzw} shows, that in many occasions perturbation theory at one-loop order finds strong transitions, while the two-loop study reveals that these transitions are in fact very weak, if exist at all.

\subsection{Parameter space scan}

\begin{figure}[t]
   \centering
   \includegraphics[width=0.4\textwidth]{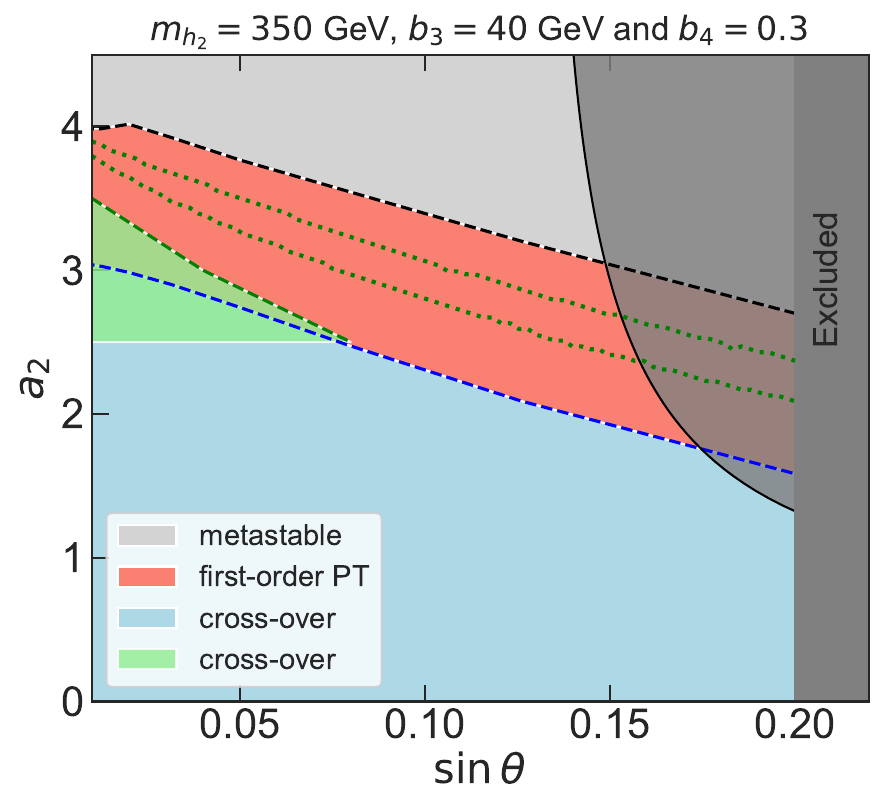}
   \includegraphics[width=0.42\textwidth]{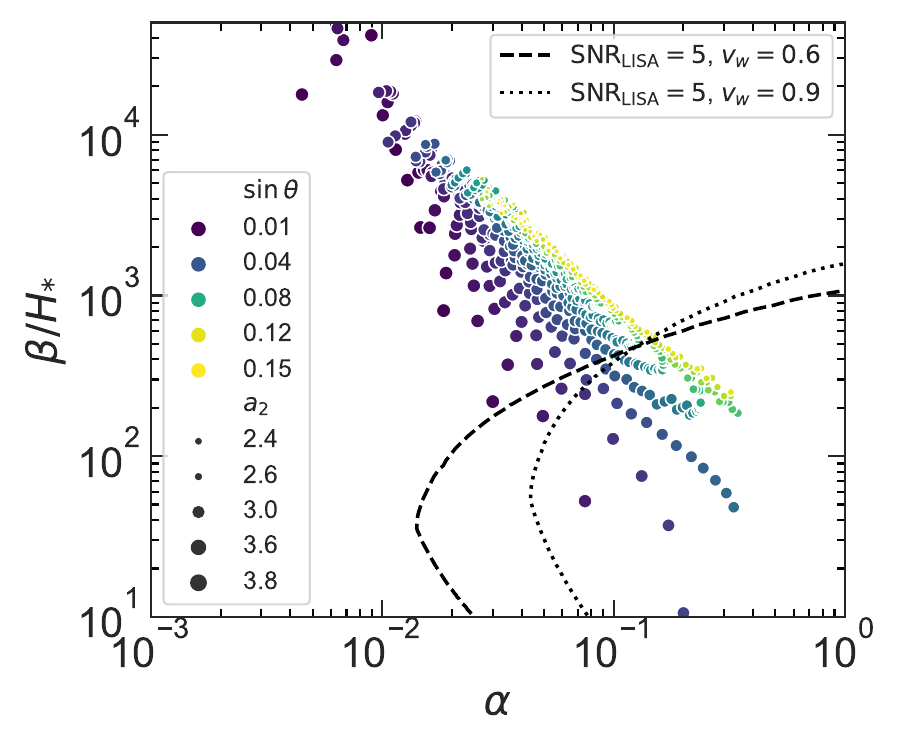}
   \includegraphics[width=0.42\textwidth]{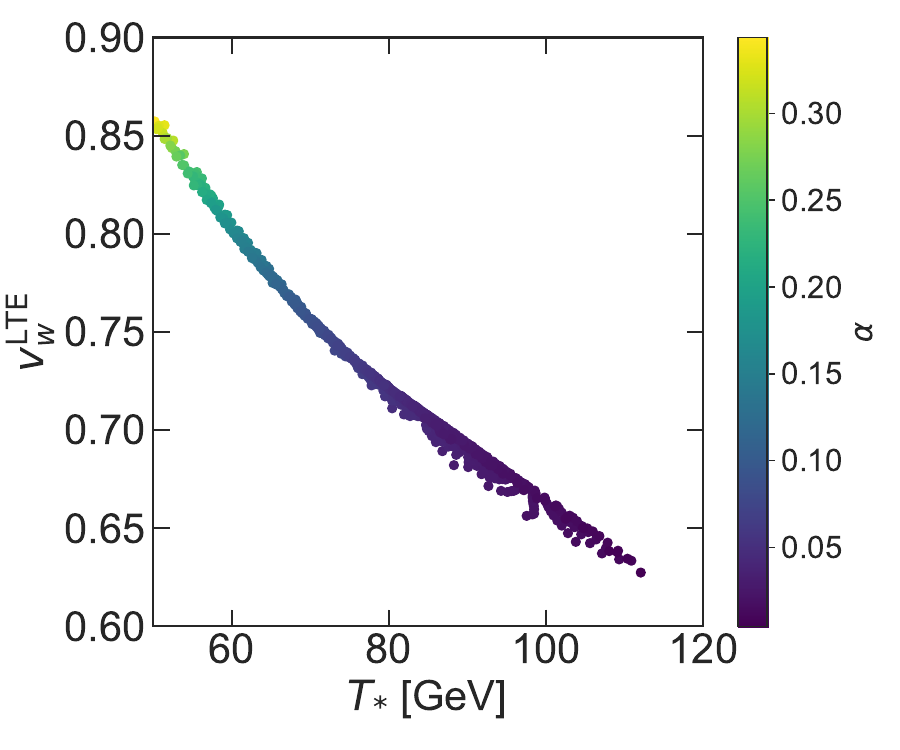}
   \includegraphics[width=0.42\textwidth]{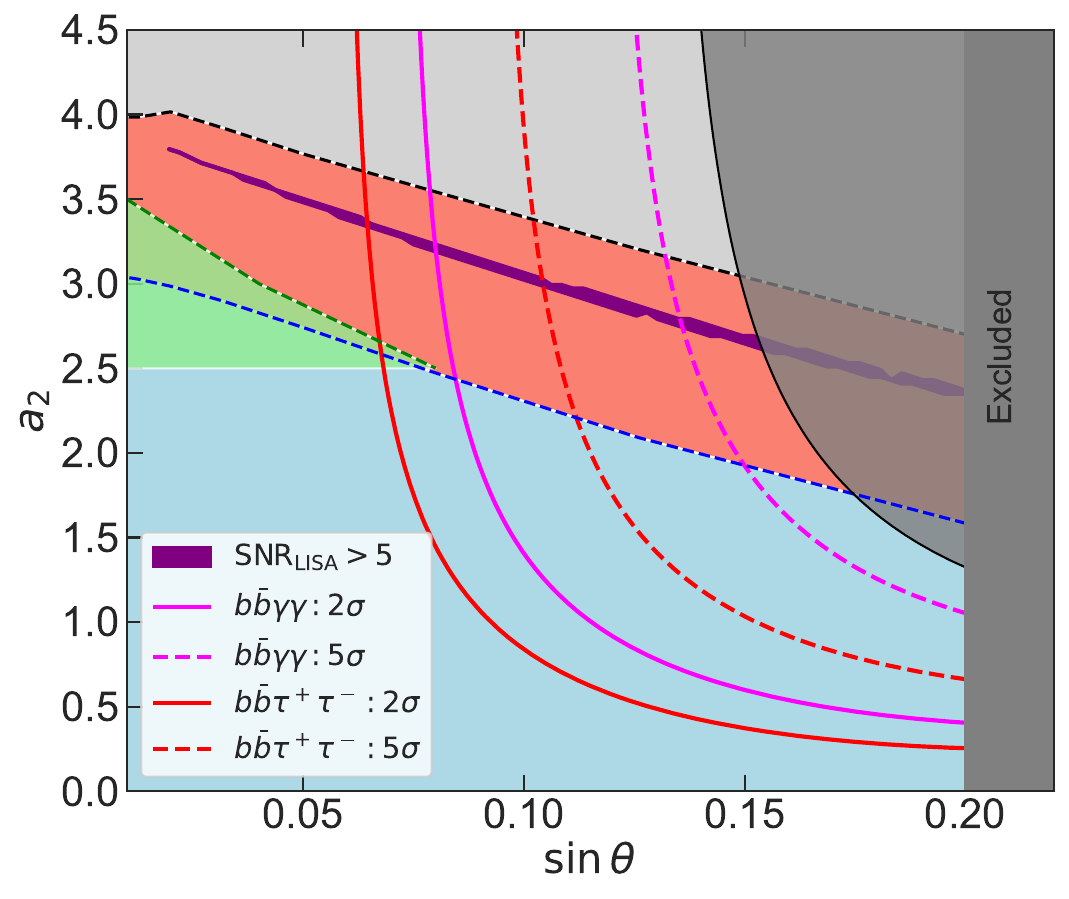}
   \caption{
    \label{fig:sintheta-a2} {\footnotesize The results for a scan over ($\sin\theta$, $a_2$) plane with fixing $m_{h_2} = 350$ GeV, $b_3 = 40$ GeV, $b_4 = 0.3$. 
{\bf Top left panel}: 
The phase structure diagram:  
In red we show the region with first-order EWPT, while blue and green regions have a cross-over, according to simulations without, and with an active singlet, respectively. The region within red between two green dotted lines denotes the region where nucleation completes. In the light gray region EW vacuum is metastable and dark gray regions are experimentally excluded. 
    {\bf Top right panel: } Scanned points in the region of completed nucleation projected on ($\alpha$, $\beta/H_*$) plane. 
    The colour hue of scatter points represents the value of $\sin\theta$ while their size indicates the value of $a_2$. 
    The dashed black and dotted black lines denote the LISA sensitivities with fixing $v_w = 0.6$ and $v_w = 0.9$ respectively. 
    {\bf Bottom left panel: } The bubble wall velocity as a function of nucleation temperature. The color represents the value of $\alpha$. 
    {\bf Bottom right panel: } The LISA sensitivity region (purple) and the significance of the di-Higgs $b\bar b \gamma\gamma$ (magenta) and $b\bar b\tau^+\tau^-$ (red) searches at the HL-LHC overlaid on the phase structure diagram. Dashed (solid) line corresponds to $5\sigma$ ($2\sigma$) significance. } 
    }
\end{figure}

\begin{figure}[t]
   \centering
   \includegraphics[width=0.42\textwidth]{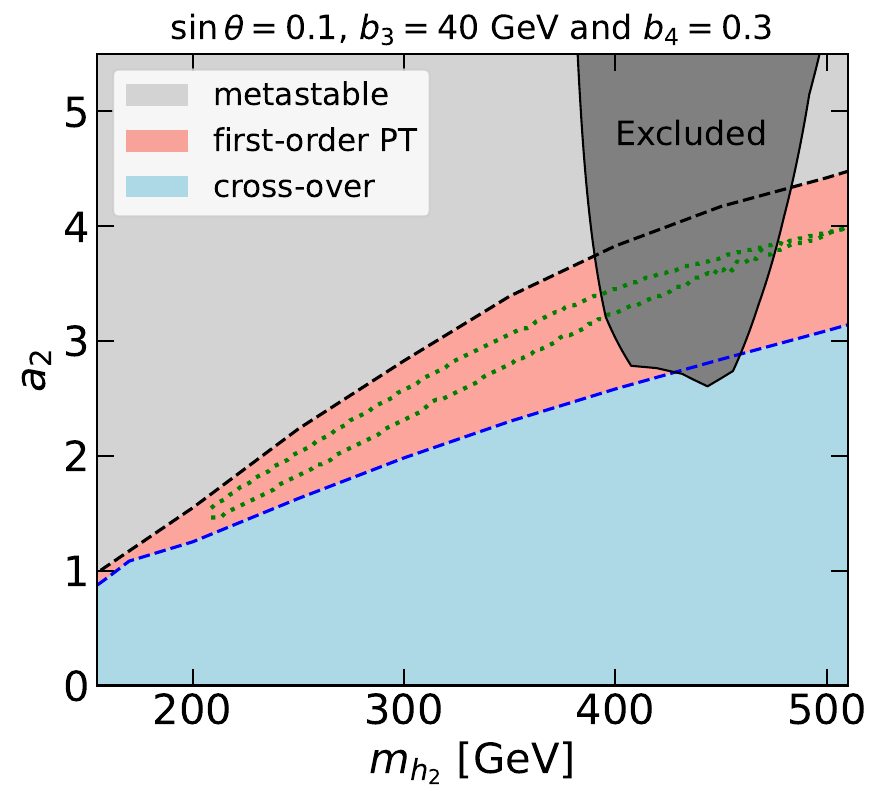}
   \includegraphics[width=0.45\textwidth]{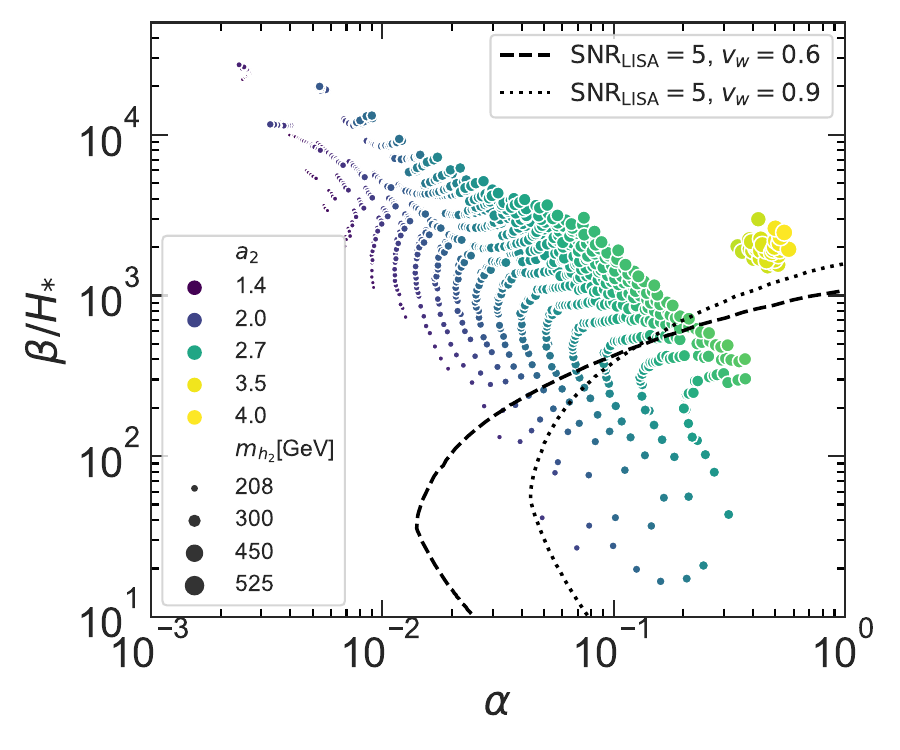}
   \includegraphics[width=0.45\textwidth]{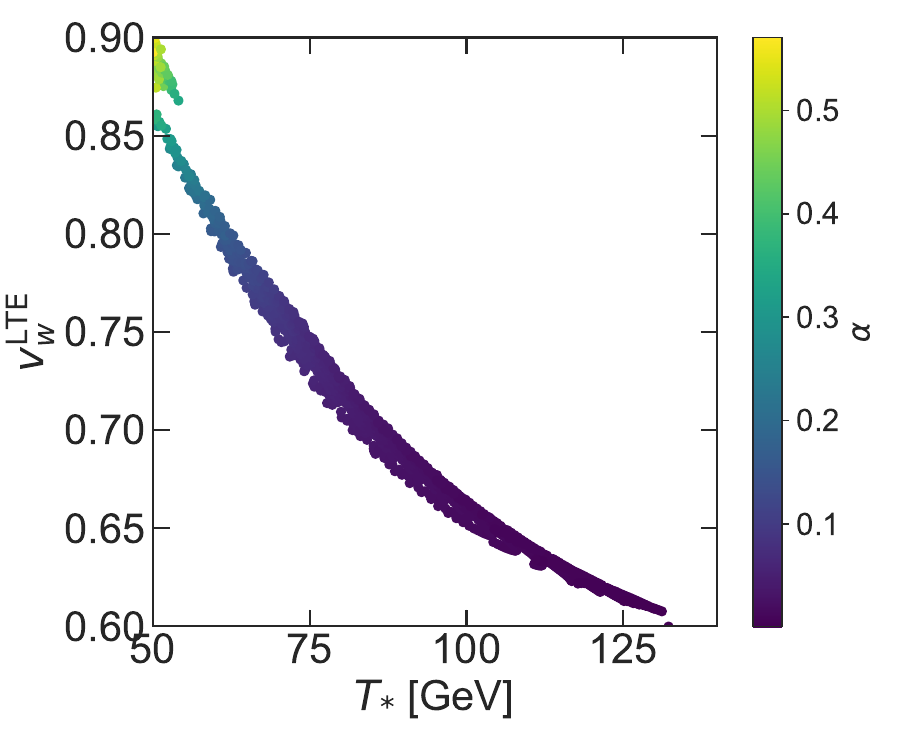}
   \includegraphics[width=0.45\textwidth]{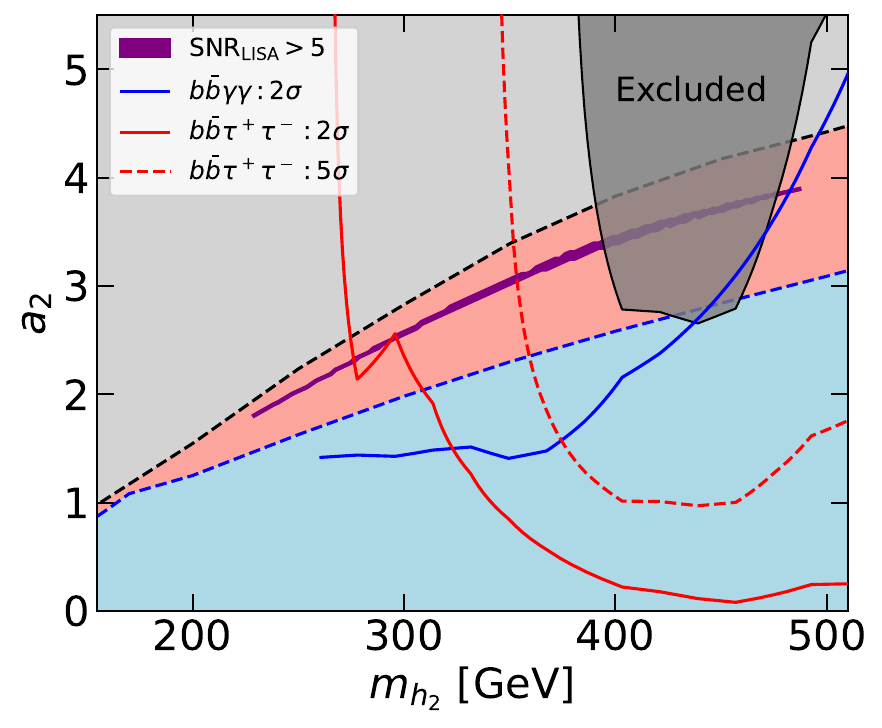}
   \caption{
    \label{fig:mS-a2}
    Similar to Fig.~\ref{fig:mS-a2} but on the ($m_{h_2}$, $a_2$) plane with fixing $\sin\theta = 0.1$, $b_3 = 40$ GeV, $b_4 = 0.3$. }
\end{figure}

Next, we perform scans over the free parameter space in the model, while incorporating two-loop thermal corrections to the scalar effective potential.
We fix singlet self-interaction couplings $b_3 = 40$ GeV and $b_4 = 0.3$, and conduct two scans across the remaining parameter space in the model: 

\begin{itemize}

\item[] 1) the first scan examines the ($\sin\theta$, $a_2$) plane with $m_{h_2}$ fixed at 350 GeV, 

\item[] 2) the second scan explores the ($m_{h_2}$, $a_2$) plane while fixing $\sin \theta = 0.1$.

\end{itemize}
Our findings from these scans are depicted in Fig. \ref{fig:sintheta-a2} and Fig. \ref{fig:mS-a2}, respectively.

The top left panel of Fig.~\ref{fig:sintheta-a2} shows the phase structure diagram on the ($\sin\theta$, $a_2$) plane. 
In the light gray region at large $a_2$, the electroweak vacuum at zero temperature is metastable, i.e. not the global minimum, and hence this region is theoretically unviable.
Dark gray regions, predominantly located at higher $\sin\theta$ values, 
are experimentally excluded
due to the current Higgs signal strength measurements 
and di-Higgs $b\bar b \tau^+ \tau^-$ searches conducted by ATLAS \cite{ATLAS:2022vkf,ATLAS-CONF-2021-030}.  

The red color region indicates the first-order phase transition occurring in the second step of the two-step transition described in (\ref{eq:PTpattern}). The blue and green regions indicate the cross-over, {\it i.e.} the lack of a phase transition. 
The dashed blue and dashed green lines represent the boundary between the cross-over 
and first-order EWPT regions, determined by utilizing 
results from the lattice simulations.  
In particular, for the dashed blue boundary line, we have assumed that the singlet is heavy enough within the intermediate EFT, enabling to integrate it out to obtain a SM-like EFT, see Appendix~\ref{sec:matching}. 
The phase diagram for such the SM-like EFT is known non-perturbatively \cite{Rummukainen:1998as}: the first-order transitions correspond to $0 < x_c < 0.11$ while $x_c > 0.11$ indicates a crossover. Here the dimensionless ratio $x_c = x(T_c) \equiv \tilde{\lambda}_{3}(T_c)/\tilde{g}^2_{3} (T_c)$ with $T_c$ is critical temperature, $\tilde{\lambda}_{3}$ is thermal Higgs self-coupling and $\tilde{g}^2_{3}$ the gauge coupling within the SM-like EFT. Hence, the dashed blue boundary line correspond to points for which $x_c = 0.11$.  

On the other hand, the dashed green boundary line is obtained by using the results from the recent lattice simulations of \cite{Niemi:2024axp} wherein the singlet scalar is kept within the final EFT and hence also actively present in the simulations. 
By comparing with the dashed blue boundary line, 
we see that the two approaches agree qualitatively, yet disagree on the exact location of the boundary for small mixing angles. 
In particular, the result of \cite{Niemi:2024axp} admits smaller first-order region for mixing angles $\sin\theta < 0.07$, 
at least up to values used in these fresh simulations with the singlet scalar.  

We find that a relatively large mixing angle can significantly influence the first-order EWPT region. Particularly, an increase in the mixing angle allows smaller value of $a_2$ to maintain the first-order EWPT, as visible in the top left panel of Fig.~\ref{fig:sintheta-a2}. 
We note that the first-order region is approximately parallel with the boundary to metastable region.
Within the same panel, we determine the region where the nucleation completes, marked by the region inside the two dotted green lines. It is worth to note, that current di-Higgs $b\bar b \tau^+ \tau^-$ searches at ATLAS constrain this region of viable nucleation to values below $\sin\theta \lesssim 0.15$.

The top right panel of Fig.~\ref{fig:sintheta-a2} illustrates the strength and duration of the EWPT, for the data points within the viable nucleation region (shown in top left panel).
We find that small changes in either the portal coupling $a_2$ or the mixing angle lead to a significant impact on both the strength and duration of the EWPT.
Specifically, higher values of either $a_2$ or the mixing angle result in larger $\alpha$ and smaller $\beta/H_{*}$, i.e. points moving towards the range of LISA sensitivity.
Within the same panel, we observe that the sensitivity region from LISA detector is influenced by the bubble wall velocity. Notably, when $v_w = 0.6$, the sensitivity line can probe a broader region characterized by small $\beta/H_{*}$ and small $\alpha$, while restricting the exploration of regions with larger $\beta/H_{*}$ and larger $\alpha$, compared to $v_w = 0.9$.

In the bottom left panel of Fig.~\ref{fig:sintheta-a2} we show the bubble wall velocity $v_w$ computed under the LTE approximation as a function of percolation temperature $T_*$. A strong correlation among $v_w$, $T_*$ and $\alpha$ is found. In particular, a larger $T_*$ results in a smaller $v_w$ and smaller $\alpha$. 
Overall, in this parameter space of interest, the bubble wall velocity varies in range of [$0.63$, $0.85$] while the percolation temperature ranges from 50 GeV to 115 GeV.%
\footnote{
The percolation temperature below 50 GeV can result in large bubble wall velocity and strong GW signal. However, due to concerns of validity arising from the high-temperature expansion considered in our analysis (c.f~\cite{Niemi:2024vzw}), we do not include these low temperature regions. 
} 

The bottom right panel of Fig.~\ref{fig:sintheta-a2} delineates the parameter space accessible within the model by both GW and HL-LHC detectors. 
Notably, the LISA sensitivity region spans in a thin band with the ranges of $0.02 < \sin\theta < 0.15$ and $2.3 < a_2 < 3.8$. The upper limit on $\sin\theta$, and consequently the imposition of a lower bound on $a_2$, within the LISA sensitivity region are due to the current constraint from di-Higgs $b\bar b \tau^+ \tau^-$ searches at ATLAS.
Furthermore, overlap between the LISA sensitivity regions and HL-LHC di-Higgs production are identified. For the HL-LHC di-Higgs $b \bar b \gamma \gamma$ search, the overlap is observed in the region $\sin\theta > 0.135$ ($\sin\theta > 0.08$) corresponding to $5\sigma$ ($2 \sigma$) significance. On the other hand, for the HL-LHC di-Higgs $b \bar b \tau^+ \tau^-$ search, a larger overlapping region is found, notably $\sin\theta > 0.1$ ($\sin\theta > 0.065$) corresponding to $5\sigma$ ($2 \sigma$) significance.

From these results, we summarise the interplay between GW and collider signals as follows:
\begin{itemize}

    \item If both LISA and HL-LHC di-Higgs searches detect the signals, the xSM model can be simultaneously responsible for both, with model parameters corresponding to the $5\sigma$ significance overlapped regions. 

    \item If LISA detects the signals but HL-LHC does not, it would exclude a large region of the parameter space with the mixing angle $\sin\theta > 0.08$ for $b \bar b \gamma \gamma$ search and $\sin\theta > 0.065$ for $b \bar b \tau^+ \tau^-$ search. However, smaller values of $\sin\theta$ could still account for the GW signal detected by LISA. Precision measurements of the Higgs boson at future collider detectors (see Ref. \cite{deBlas:2019rxi}) could further probe these lower values of the mixing angle and validate the GW signals. 

    \item Conversely, if the HL-LHC detects the signals but LISA does not, the xSM model could account for the HL-LHC signals  but a narrow band on ($\sin\theta$, $a_2$) plane addressed to LISA would be excluded.  

\end{itemize}

Fig.~\ref{fig:mS-a2} illustrates results similar to to Fig.~\ref{fig:sintheta-a2}, but in the ($m_{h_2}$, $a_2$) plane with fixed $\sin \theta = 0.1$, $b_3 = 40$ GeV and $b_4 = 0.3$. However, a boundary line between the crossover and the first-order EWPT regions in the case of the dynamical singlet is not presented, due to the absence of the lattice results in this case.

We find that as the mass of the singlet-like state $h_2$ increases, larger values of the parameter $a_2$ are necessary to accommodate the first-order EWPT. Scanning across the first-order EWPT region, we identify a region of viable nucleation 
between the green dotted lines in the top left panel of Fig.~\ref{fig:mS-a2}, for which $200$ GeV $< m_{h_2} < 520$ GeV. These bounds on the new scalar mass are more stringent than those derived from
the requirement of a first-order phase transition as shown by the red region (see also in \cite{Ramsey-Musolf:2024zex}). 
Moreover, it is worth noting that a segment of this nucleation-viable region, specifically $395$ GeV $< m_{h_2} < 475$ GeV, is excluded by the current constraint from di-Higgs $b\bar b \tau^+\tau^-$ at ATLAS.

Analysis of the top right panel of Fig.~\ref{fig:mS-a2} reveals a notable trend: heavier $h_2$ masses and broader $a_2$ regions correspond to increased values of $\beta/H_{}$ and $\alpha$. Note that a gap between the distinct points observed in the higher $\beta/H_{*}$ and $\alpha$ ranges and the remainder is a consequence of the current constraint from di-Higgs $b\bar b \tau^+\tau^-$ at ATLAS. 

Similar to our findings in Fig.~\ref{fig:sintheta-a2}, we observe a strong correlation among the bubble wall velocity, nucleation temperature, and the strength of the phase transition, depicted in the bottom left panel of Fig.~\ref{fig:mS-a2}. The bubble wall velocity varies within the range of $0.6$ to $0.9$, while the nucleation temperature extends from $50$ GeV to $130$ GeV.

Finally, we show the parameter space within the ($m_{h_2}$, $a_2$) plane conducive to detectable GW signals at LISA and di-Higgs production signals at HL-LHC detectors. The region probed by LISA spans on the new scalar mass from $230$ GeV to $395$ GeV and a smaller segment observed at mass from $475$ GeV to $485$ GeV. We obtain the overlapping regions between LISA sensitivity and HL-LHC di-Higgs production searches. Particularly, the overlapping regions for the $b\bar{b}\tau^+\tau^-$ search, which manifest for masses $m_{h_2} > 350$ GeV and $m_{h_2} > 275$ GeV, achieving $5\sigma$ and $2\sigma$ significance, respectively. 
On the other hand, the $b\bar{b}\gamma\gamma$ search targets a lower mass range, notably $m_{h_2} > 260$ GeV for $2\sigma$ significance. However, it does not encompass the LISA sensitivity regions for $5\sigma$ significance.

In analogy to the findings from our previous scan above, we summarise:
\begin{itemize}
    \item If signals observed in both the LISA detector and the HL-LHC di-Higgs $b\bar{b}\tau^+\tau^-$ search, the xSM model can concurrently account for both observations with $m_{h_2} > 350$ GeV and the coupling $a_2$ falls within the range [$3$, $4$], assuming $\sin\theta = 0.1$. However, should signals be solely detected in the HL-LHC di-Higgs $b\bar{b}\gamma \gamma$ search, a larger $\sin\theta$ may be necessary, as suggested by the previous scanning results from above.

    \item In scenarios where LISA detects signals but HL-LHC does not, exclusion of the heavy mass region occurs. Specifically, exclusion criteria entail $m_{h_2} > 275$ GeV for the $b\bar{b}\tau^+\tau^-$ search and $m_{h_2} > 260$ GeV for the $b\bar{b}\gamma\gamma$ search. Nevertheless, a smaller mass region accommodating GW signals detected by LISA remains viable.

    \item On the other hand, if HL-LHC detects signals but LISA does not, the xSM model may elucidate the HL-LHC signals within a heavy scalar mass region. However, this would entail exclusion of a narrow band within the ($m_{h_2}$, $a_2$) plane associated with LISA detections. 

\end{itemize}

\section{Discussion}
\label{sec:conclusion}

In this article, we have performed a cutting-edge analysis of GW signals stemming from the first-order EWPT, while concurrently exploring their interplay with collider phenomenology within the framework of the scalar singlet extension of the SM. Our main results are shown in Figs.~\ref{fig:sintheta-a2} and~\ref{fig:mS-a2}. 
For collider arena, we focused on signals from di-Higgs production, specifically targeting the $b\bar{b}\tau^+ \tau^-$ and $b\bar{b}\gamma\gamma$ final states at HL-LHC. 
The search for the di-Higgs decay into $b\bar b\tau^+\tau^-$ final state at the HL-LHC turned out to be more sensitive to heavy singlet mass regions while the search for $b\bar b\gamma \gamma$ final state is more sensitive to lighter mass regions. 

For the GW predictions, our analysis employs and further develops state-of-the-art techniques, including the use of dimensionally reduced
effective field theory to describe thermodynamics of the primordial plasma.
By using thermal effective potential with several two-loop corrections at high-temperature expansion, we can achieve a significant reduction in uncertainties regarding
thermal parameters for GW predictions \cite{Croon:2020cgk,Gould:2021oba,Niemi:2024vzw}, also bringing our results in closer alignment with results from lattice simulations \cite{Niemi:2024axp}. Furthermore, we have employed $\hbar$-expansion for the effective potential to ensure gauge invariance and perturbative consistency in our results. Additionally, we demonstrated the possibility to utilize non-perturbative simulation results to determine phase structure diagrams of the xSM, and prove the existence of first-order phase transitions. 

For the first time in the context of dimensionally reduced EFTs, we estimated the bubble wall velocity using the local thermal equilibrium approximation. 
With this upgrade, for the first time we have computed all four thermal parameters $(T_*,\alpha_*,\beta/H_*, v_w)$ entering the determination for GW spectrum, while including several two-loop level thermal effects. For other three thermal parameters these two-loop effects are known to be crucially important -- due to slow convergence of perturbation theory at high temperatures \cite{Gould:2021oba} -- in order to reach reliable results. 
For the bubble wall speed (in LTE approximation) we found that higher order thermal corrections in many parameter points with strong transitions lead to increasing result, due to a positive correlation of the bubble wall speed and the phase transition strength $\alpha$. We also observed a strong correlation among the bubble wall velocity, nucleation temperature, and the strength of the phase transition.

Interestingly the GW and collider signals can be simultaneously detectable in the region where the new heavy scalar boson mass lies in specific ranges (see Sec.~\ref{sec:intro}) below $\sim 500$ GeV and the mixing angle between the Higgs and new scalar boson is relatively large. 
We note that, we have only focused on the heavy mass range, i.e. the new scalar boson being heavier than the SM Higgs. We defer the light mass scenario $m_{h_2} < 125.1$ GeV to a future study.

\begin{acknowledgments}
We would like to thank Benoit Laurent, Jorinde van de Vis and Jiang Zhu for enlightening discussions regarding the computation of the bubble wall velocity, and Lauri Niemi and Guotao Xia for useful discussions on their lattice simulation results, as well as Oliver Gould and Paul Saffin for discussions related to their upcoming work, similar to ours. 
This work was supported in part by the National Natural Science Foundation of China, grant Nos. 19Z103010239 and 12350410369 (VQT).
VQT would like to thank the Medium and High Energy Physics group at the Institute of Physics, Academia Sinica, Taiwan for their hospitality during the course of this work.
\end{acknowledgments}

%\newpage
\appendix
\allowdisplaybreaks
%\begin{center}{\large \textbf{Appendix}} \end{center}

\section{Experimental constraints}
\label{app:constraints}

The extension of the scalar sector in the model can be constrained by the data from Higgs search experiments and the measurements of the SM-like Higgs boson at the LHC. The constraints are placed on the mixing angle $\theta$ and the mass of the extra scalar boson.  This is similar to that of dark doublet Higgs extension of the SM as studied in \cite{Tran:2023lzv}.

The couplings of the physical scalars $h_1$ and $h_2$ to the SM gauge boson and fermions can be given as 
\be
{\mathcal L}_{\rm Higgs} \supset  \frac{h_1 \cos \theta - h_2 \sin \theta}{v} \left(2 m_{W}^2 W^{+}_\mu W^{-\mu} + m_Z Z_\mu Z^\mu - \sum_{f} m_f {\bar{f}} f\right) \; . 
\ee
While the SM-like Higgs boson $h_1$ couples to the SM particles are modified by a factor of $\cos\theta$, the heavier Higgs $h_2$ couples to them with a suppression factor of $(-\sin\theta)$.
The Higgs boson signal strength can then by given by 
\be
\label{eq:muh1}
\mu_{h_1} \equiv \cos^2\theta \frac{{\rm BR}(h_1 \to {\rm SM})}{{\rm BR^{SM}}(h_1 \to {\rm SM})} \; ,
\ee
where $\text{BR}^{\text{SM}}(h_1\rightarrow \text{SM}) \equiv 1$ and $\text{BR}(h_1\rightarrow \text{SM}) = \frac{ \Gamma_{h_1}^{\text{SM}}\cos^2\theta}{ \Gamma_{h_1}^{\text{SM}}\cos^2\theta +\Gamma_{h_1\to h_2h_2} }$ with $\Gamma_{h_1\to h_2h_2}$ is the partial decay width of $h_1 \to h_2 h_2$. In this analysis, we consider $m_{h_2} > 2 m_{h_1}$ hence the decay of $h_1 \to h_2 h_2$ is kinematically forbidden. Therefore, the Higgs boson signal strength in \ref{eq:muh1} becomes $\mu_{h_1} = \cos^2\theta$.
Using the current combined Higgs signal strengths measurement from ATLAS~\cite{ATLAS:2022vkf}
\begin{equation}
\mu_{h_1} = 1.05 \pm 0.06  \; ,
\end{equation}
one can obtain a bound on the mixing angle $|\sin\alpha |\lesssim 0.2$ at $95\%$ C.L. 

\begin{figure}[tb]
\centering
\includegraphics[width=0.5\textwidth]{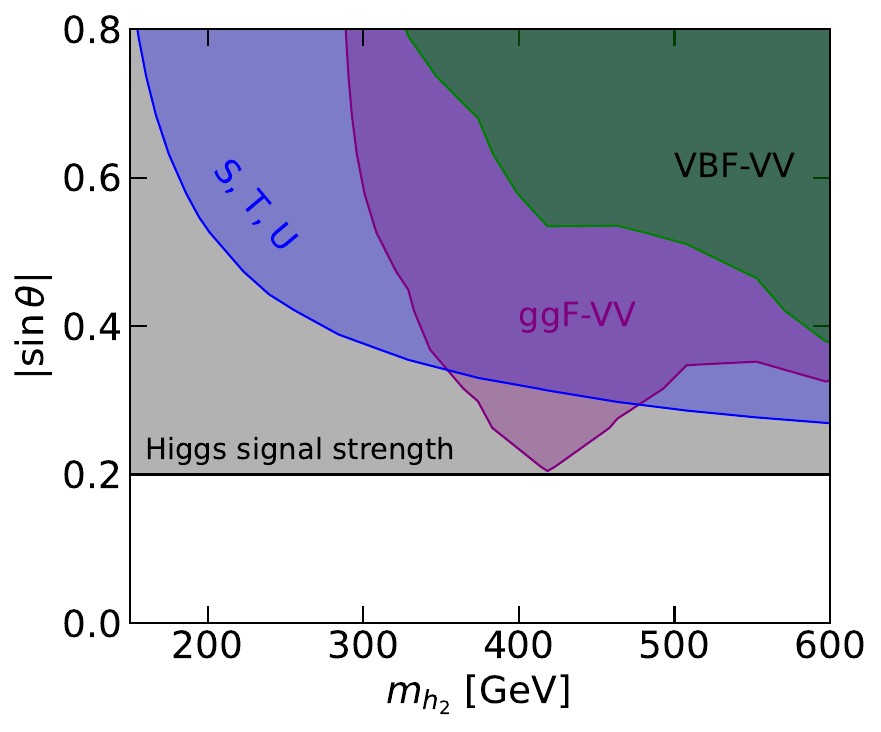}
\caption{\label{fig:constraint}
Upper bounds on the mixing angle $\theta$ as a function of the heavy Higgs mass $m_{h_2}$. The color shaded regions represent the exclusion regions from the oblique parameter constraint (blue region), the di-boson searches via gluon-gluon fusion (purple region) and vector boson fusion (green region) channels at ATLAS~\cite{ATLAS:2020fry}, and the combined Higgs signal strength (gray region) at ATLAS~\cite{ATLAS:2022vkf}.
    }
\end{figure}

The current direct heavy resonance searches at the LHC can put constraints on the mass of new scalar and its mixing angle to Higgs boson. 
Here we utilize the measurements on the heavy diboson resonances in semileptonic final states data at ATLAS \cite{ATLAS:2020fry}. 
The constraints on ($m_{h_2}$, $|\sin\theta|$) plane from the heavy diboson resonances are depicted as purple and green shaded regions in the right panel of Fig.~\ref{fig:constraint}. 

The oblique parameters $S$, $T$, and $U$ \cite{Peskin:1991sw} can be modified in xSM. 
Particularly, the oblique parameter $\mathcal O$ can be given as \cite{Profumo:2014opa} 
\be
\Delta {\mathcal O} = \left[{\mathcal O}^{\rm SM} (m_{h_2}) - {\mathcal O}^{\rm SM} (m_{h_1}) \right] \, \sin^2 \phi \; ,
\ee
where ${\mathcal O}^{\rm SM}$ is the oblique parameter given in the SM. 
We use the global fit values for the oblique parameters at 
Particle Data Group (PDG)~\cite{ParticleDataGroup:2020ssz},
which are given as 
\bea
\label{eq:STU_pdg}
\Delta S &=& -0.01 \pm 0.1 \; , \nonumber \\
\Delta T &=& 0.03 \pm 0.12 \; ,  \\
\Delta U &=& 0.02 \pm 0.11 \; , \nonumber
\eea 
and the correlation coefficients are $0.92, -0.8$ and $-0.93$ for ($\Delta S, \Delta T$), ($\Delta S, \Delta U$) and ($\Delta T, \Delta U$), respectively. 
The constraint on ($m_{h_2}$, $|\sin\theta|$) plane from the oblique parameters are shown as the blue shaded region in the right panel of Fig.~\ref{fig:constraint}. One can see that the upper bound on the mixing angle becomes more stringent in the heavier mass region of $h_2$.

\section{Matching relations for thermal EFTs}
\label{sec:matching}

In this appendix, we collect the matching relations between parameters of the full parent theory (xSM) and its effective theories at high temperatures, at different thermal scales. Most of the results in this section were originally obtained in \cite{Kajantie:1995dw,Niemi:2021qvp,Schicho:2021gca} and can also be obtained using \texttt{DRalgo} package \cite{Ekstedt:2022bff}. 

\subsection{Integrating out the non-zero Matsubara modes}

The dimensional reduction from full theory at four dimensions to thermal EFT at three dimensions proceeds by integrating out the non-zero Matsubara modes.  
Consequently, the Euclidean action in three-dimensional EFT for the xSM reads 
\begin{align}
\label{eq:3d-action}
S_\text{3d} =& \int d^3 x \Big\{
  \frac14 F^a_{ij} F^a_{ij}
+ \frac14 B_{ij} B_{ij} 
+ |D_i \phi|^2
+ \frac12 (\partial_i S)^2
+ V^{\text{3d}}(\phi, S )
\nonumber \\[1mm]
&
+ \frac12 (D_i A_0^a)(D_i A_0^a)
+ \frac12 m_D^2 A^a_0 A^a_0
+ \frac12 (\partial_i B_0)^2  
+ \frac12 (m_D')^2 B_0^2
\nonumber \\[1mm]
&
+ \frac12 (D_i C_0^\alpha)(D_i C_0^\alpha)
+ \frac12 (m_D'')^2 C_0^\alpha C_0^\alpha
+ h_3 \phi^\dagger \phi A_0^a A_0^a + h_3' \phi^\dagger \phi B_0^2
+ h_3'' \phi^\dagger A_0^a \sigma_a \phi B_0 \nonumber \\[2mm]
&
+ \omega_3 \phi^\dagger \phi C_0^\alpha C_0^\alpha
+ x_3 S A_0^a A_0^a
+ x_3' S B_0^2
+ y_3 S^2 A_0^a A_0^a
+ y'_3 S^2 B_0^2
\nonumber \\[2mm]
&
+  \text{interactions among } {A_0, B_0} \text{ and } C_0 \Big\}.
\end{align}
Here $\sigma_a$ are the Pauli matrices with isospin index $a=1,2,3$ and $F_{ij}, B_{ij}$ are field strength tensors (with spatial Lorentz indices $i,j=1,2,3$) for the SU(2) and U(1)$_Y$ gauge fields whose couplings are denoted by $g_3$ and $g_3'$. The temporal gauge field components $A^a_0, C^\alpha_0$ are Lorentz scalars in adjoint representations of SU(2) and SU(3), respectively (with adjoint colour index $\alpha = 1,...,8$), while $B_0$ scalar is U(1)$_Y$ singlet. 
The scalar potential in \eqref{eq:3d-action} is given by 
\begin{align}
\label{eq:V3d}
V^{\text{3d}}(\phi,S) &=   \frac12 {\mu}^2_{3} \phi^\dagger \phi + \frac14 {\lambda}_3 (\phi^\dagger \phi)^2  + \frac14 {a}_{1,3} \phi^\dagger \phi S + \frac14 {a}_{2,3} \phi^\dagger \phi S^2 \nonumber \\
& + {b}_{1,3} S + \frac12 {b}_{2,3} S^2 + \frac13 {b}_{3,3} S^3 + \frac14 \bar{b}_{4,3} S^4.
\end{align}
We do not use distinguished notation for three-dimensional fields $\phi$ and $S$, but note that within the EFT they have mass dimension 1/2 instead of unit mass dimension in the parent theory.  

Matching relations for the effective parameters in \eqref{eq:V3d} read%
\footnote{
Couplings $g$, $g'$, $g_s$ and $y_t$ are gauge couplings for SU(2), U(1)$_Y$ and SU(3), and top quark Yukawa coupling, respectively. 
}
\begin{align}
{\mu}_{3}^2 &= \mu_{3,\rm SM}^2  + \frac{T^2}{24} a_2 - \frac{L_b}{2 (4\pi)^2} \left( a_2 b_2 + \frac12 a_1^2 \right)  \\ \nonumber
&  + \frac{a_2 T^2}{(4\pi)^2} \left[ \frac{1}{24} L_b \left( \frac34 (3g^2 + {g'}^2) - 6 \lambda - 5 a_2 - 3 b_4 \right)  - \frac18 y_t^2 L_f - \frac{1}{2} a_{2} \left( c + \log \left( \frac{3 T}{\Lambda_3} \right) \right)  \right] \\
 {\lambda}_3 &= T \lambda + \frac{T}{(4\pi)^2} \bigg[ \frac{2-3L_b}{16} \left( 3g^4 + 2g^2{g'}^2 + {g'}^4 \right) + 3 y_t^2 L_f \left( y_t^2 - 2 \lambda \right) \\\nonumber 
 &+ L_b \left(\frac{3}{2}(3g^2 + {g'}^2) \lambda - 12 \lambda^2 - \frac{1}{4} a_2^2 \right) \bigg] 
 \\ 
{a}_{1,3} &= \sqrt{T} a_1
  + \frac{\sqrt{T}}{(4\pi)^2} \left[ L_b \left(\frac{3}{4} (3g^2 + {g'}^2) a_1 -2 b_3 a_2 - ( 6\lambda + 2a_2) a_1 \right) -3 L_f y_t^2 a_1 \right]  
  \\
{a}_{2,3} &= T a_2
  + \frac{T}{(4\pi)^2} \left[ L_b \left(\frac{3}{4}(3g^2 + {g'}^2) - 6\lambda - 2 a_2 - 3 b_4 \right)a_2  -3 L_f y_t^2 a_2 \right]
\\
{b}_{1,3} &= \frac{1}{\sqrt{T}} \bigg[b_1 + \frac{T^2}{12} \Big( b_3 + a_1 \Big) - \frac{L_b}{(4\pi)^2} \Big( a_1 \mu^2 + b_3 b_2 \Big)
  \\ \nonumber &
  + \frac{T^2}{(4\pi)^2}\bigg[ \frac{2+3L_b}{48}(3g^2 + {g'}^2) a_1 - \frac{L_b}{2} \bigg( \Big( \lambda + \frac{7}{12} a_2 \Big) a_1 + \Big( \frac{1}{3} a_2 + \frac{3}{2}b_4 \Big)b_3 \bigg) - \frac{1}{8} a_1 y_t^{2} \Big(3L_b - L_f \Big) \bigg] \bigg] \\ \nonumber
& - \frac{1}{(4\pi)^2} \bigg[ 2 {b}_{3,3} {b}_{4,3}- \frac12 {a}_{1,3} \Big(3g^2_3 + g'^2_3 - 2 {a}_{2,3} \Big) \bigg] \Big( c + \ln\Big(\frac{3T}{{\Lambda}_3} \Big) \Big) 
\\
{b}_{2,3} &=
  b_2
  + T^2 \Big(\frac{1}{6}a_2+ \frac{1}{4}b_4 \Big)
  - \frac{L_b}{(4\pi)^2}\Big( 2b_3^2+ \frac{1}{2} a_1^2 + 2 a_2 \mu^2 + 3 b_4 b_2 \Big)
  \\ \nonumber &
  + \frac{T^2}{(4\pi)^2}\bigg[
      \frac{2+3L_b}{24}(3g^2 + {g'}^2) a_2
  - L_b \bigg( \Big(\lambda + \frac{7}{12}a_2 + \frac{1}{2} b_4 \Big) a_2 + \frac94 b_4^2 \bigg)
  - \frac{1}{4} a_2 y_t^{2}(3L_b - L_f) \bigg]
\\ \nonumber
& + \frac{1}{(4\pi)^2} \Big( (3g^2_3 + g'^2_3) {a}_{2,3} - 2 {a}_{2,3}^2 - 6 {b}_{4,3}^2 \Big) \Big( c + \ln\Big(\frac{3T}{\Lambda_3} \Big) \Big) .
\\
{b}_{3,3} &= \sqrt{T} b_3
  - \frac{\sqrt{T}}{(4\pi)^2} 3 L_b \Big( \frac{1}{2} a_1 a_2 + 3 b_4 b_3 \Big)
\\
{b}_{4,3} &= T b_4
  - \frac{T}{(4\pi)^2} L_b \Big( a_2^2 + 9 b_4^2 \Big), 
\end{align}
where the SM part $\mu^2_{3,\rm SM}$ can be found in Refs.~\cite{Kajantie:1995dw,Schicho:2021gca} and 
\bea
L_b & =&  2\log \left(\frac{\Lambda}{T}\right) - 2 \left[ \log (4\pi) - \gamma \right], \\
L_f &=& L_b + 4 \log(2), \\
c &=& \frac{1}{2} \left[ \log(8\pi/9)  + \frac{\zeta'(2)}{\zeta(2)} - 2 \gamma \right]
\eea
Here, $\Lambda$ and $\Lambda_3$ are renormalization scales in the parent theory and the EFT, respectively, in the minimal subtraction scheme. 

Matching relations for the SU(2) and U(1) gauge couplings read
\begin{align}
\nonumber g_3^2={}&g^2 T\bigg[1 +\frac{g^2}{(4\pi)^2}\bigg(\frac{44-N_d }{6}L_b
+\frac{2}{3}-\frac{4N_f}{3}L_f\bigg)\bigg],\\
g'^2_3={}&g'^2T\bigg[1 +\frac{g'^2}{(4\pi)^2}\bigg(-\frac{N_d}{6}L_b-\frac{20N_f}{9}L_f\bigg)\bigg]
\end{align}

Finally, the matching relations for the Debye masses of temporal gauge field components, and couplings between them and doublet and singlet scalars in \eqref{eq:3d-action} read
\bea
m_D^2 &=& g^2 T^2 \left( \frac{4+N_d} {6} + \frac{N_f}{3} \right), \\
m_D'^2 &=&  g'^2 T^2 \left( \frac{N_d}{6} + \frac{5N_f}{9} \right), \\
m_D''^2 &=& g_s^2 T^2 \left( 1 + \frac{N_f}{{3} } \right), 
\\
x_3 &=& \frac{{ \sqrt{T} }}{(4\pi)^2} g^2 a_1,
\\
x_3' &=& \frac{{ \sqrt{T} }}{(4\pi)^2} {g'}^2 a_1,
\\
y_3 &=& \frac{T}{(4\pi)^2} \frac{1}{2} g^2 a_2, 
\\
y_3' &=&  \frac{T}{(4\pi)^2} \frac{1}{2} {g'}^2 a_2, 
\\
\notag
h_3  &=& \frac{g^2 T}{4}\bigg(1+\frac{1}{(4\pi)^2}\bigg\{\bigg[\frac{44-N_d}{6}L_b+\frac{53}{6}-\frac{N_d}{3} -\frac{4N_f}{3}(L_f-1)\bigg]g^2 \\
&& +\frac{g'^2}{2} -6 y_t^2 + 12\lambda \bigg\} \bigg), 
\\
h'_3 &=& \frac{g'^2T}{4}\bigg(1 +\frac{1}{(4\pi)^2}\bigg\{\frac{3g^2}{2}+\bigg[\frac{1}{2}-\frac{N_d}{6}\Big(2+L_b \Big)  -\frac{20N_f}{9}(L_f-1)\bigg]g'^2 \nonumber 
\\
&&- \frac{34}{3} y_t^2 + 12\lambda \bigg\} \bigg),
\\
h''_{3}&=&\frac{g g' T}{2}\bigg\{1+\frac{1}{(4\pi)^2}\bigg[-\frac{5+N_d}{6} g^2+ \frac{3-N_d}{6}g'^2+L_b\bigg(\frac{44-N_d}{12}g^2 -\frac{N_d}{12}g'^2\bigg) \nonumber 
\\
&&-N_f(L_f-1)\bigg(\frac{2}{3}g^2+\frac{10}{9}g'^2\bigg) + 2 y_t^2 + 4 \lambda \bigg]\bigg\},
\\
\omega_3 &=& -\frac{2 T}{16 \pi^2} g^2_s y_t^2, \\
\eea
where $N_d = 1$ is the number of Higgs doublets and $N_f = 3$ is the number of fermion generations.

\subsection{Integrating out the temporal gauge field components}

Next, we integrate out the temporal gauge field components $A^a_0$, $B_0$ and $C^\alpha_0$ which are characterized by the Debye mass scale of $gT$. 
The matching relations at this scale are given by 
\bea
\bar{b}_{1,3} &=& b_{1,3} - \frac{1}{4\pi} \Big( 3 m_D x_3 + m_D' x_3' \Big), \\
\bar{b}_{2,3} &=& b_{2,3} - \frac{1}{2\pi} \Big( 3m_D y_3 + m_D' y_3' + \frac{3x_3^2}{2 m_D} + \frac{{x'_3}^2}{2 m_D'} \Big), \\
\bar{\lambda}_3 &=& \lambda_3 - \frac{1}{2(4\pi)} \Big( \frac{3h_3^2}{m_D} + \frac{{h'_3}^2}{m_D'} + \frac{{h''_3}^2}{m_D + m_D'} \Big), \\ 
\bar{g}_3^2 &=& g_3^2 \left(1 - \frac{g_3^2}{6(4\pi) m_D} \right), \\
\label{eq:mu3}
\bar{\mu}_{3}^2 &=& \mu_3^2 - \frac{1}{4\pi} \Big( 3 h_3 m_D + h_3' m_D' + 8 \omega_3 m_D'' \Big), \nonumber \\
 && + \frac{1}{(4\pi)^2} \Bigg[3g_3^2 h_3 - 3 h_3^2 - h_3'^2 - \frac{3}{2} h_3''^2 + \left( -\frac{3}{4} g_3^4 + 12 g_3^2 h_3 - 6 h_3^2 \right) \log\left(\frac{\Lambda_3}{2m_D}\right) 
 \nonumber \\
 &&- 2h_3'^2 \log\left(\frac{\Lambda_3}{2m_D'}\right) - 3h_3''^2 \log\left(\frac{\Lambda_3}{m_D + m_D'}\right) \Bigg].
\eea 
Working formally at ${\cal O} (g^4)$ \cite{Schicho:2021gca},  the remaining parameters $\bar{a}_{1,3}$, $\bar{a}_{2,3}$, $\bar{b}_{3,3}$, $\bar{b}_{4,3}$ and $\bar{g}'_{3,3}$ remain the same as ${a}_{1,3}$, ${a}_{2,3}$, ${b}_{3,3}$, ${b}_{4,3}$ and ${g}'_3$, respectively. 

\subsection{Integrating out the singlet scalar boson}

If the singlet is further integrated out, the final effective theory is the SM-like EFT \cite{Kajantie:1995dw}, with 
the matching relations \cite{Ekstedt:2024etx}
\bea
\label{eq:matchingSingletout_g}
\tilde{g}_3  &=& \bar{g}_3,
\\
\tilde{g}'_3  &=& \bar{g}'_3,
\\
\label{eq:matchingSingletout_lambda}
\tilde{\lambda}_3  &=& {\bar{\lambda}}_3  - \frac{a_{1,3}^2}{8 b_{2,3}} + \frac14 b_{1,3} \Big( 2\frac{a_{2,3}a_{1,3}}{b^2_{2,3}} - \frac{b_{3,3} a^2_{1,3}}{b^3_{2,3}} \Big) - \frac{1}{32\pi} \frac{a^2_{2,3}}{\sqrt{b_{2,3}}} \nonumber \\
&+& {\frac{a^2_{1,3}}{32\pi b^{\frac{3}{2}}_{2,3}}} {\bigg( 5 a_{2,3} - 12 \lambda_{3} - 3 b_{4,3} - 2 a_{2,3} \frac{b_{3,3}}{a_{1,3}} \bigg)} \nonumber \\
&+& {\frac{a^2_{1,3}}{32\pi b^{\frac{5}{2}}_{2,3}}} {\bigg( \frac{5}{4} a^2_{1,3} - a_{1,3} b_{3,3} - b^2_{3,3}   \bigg)} , 
\\
\label{eq:musingletout}
\tilde{\mu}^2_3  &=& \bar{\mu}^2_3 - \frac{a_{1,3} b_{1,3}}{2 b_{2,3}} 
    -\frac{1}{16 \pi} \left( 2 a_{2,3} \sqrt{b_{2,3}} + \frac{a_{1,3}}{\sqrt{b_{2,3}}}{(a_{1,3} - 2 b_{3,3})} \right).
\eea
We note that two-loop order corrections in \eqref{eq:matchingSingletout_lambda} and \eqref{eq:musingletout} have been neglected.

%%%%%%%%%%%%%%%%%%%%%%%%% BIBLIO %%%%%%%%%%%%%%%%%%%%%%%%%%%%%%%%%%%%%%%%%
%

\allowdisplaybreaks

\bibliographystyle{apsrev4-1}
\bibliography{refs}

\end{document}